\documentclass[sigconf]{acmart} 

\usepackage{multirow}
\usepackage{CJKutf8}
  
\AtBeginDocument{%
  }

\setcopyright{acmlicensed}
\copyrightyear{2018}
\acmYear{2018}
\acmDOI{XXXXXXX.XXXXXXX}
\acmConference[Conference acronym 'XX]{Make sure to enter the correct
  conference title from your rights confirmation email}{June 03--05,
  2018}{Woodstock, NY}

\acmISBN{978-1-4503-XXXX-X/2018/06}
\usepackage{pifont}
\usepackage{caption}
\usepackage{tabularx} 

\usepackage{subcaption}
\usepackage{enumitem}

\begin{document}

\title{Who Decides How Knowing Becomes Doing? Redistributing Authority in Human–AI Music Co-Creation}

\author{Zhejing Hu}
\email{zhejing.hu@connect.polyu.hk}
\affiliation{%
  \institution{The Hong Kong Polytechnic University}
  \city{Hong Kong}
  \country{China}
}

\author{Yan Liu}
\email{yan.liu@polyu.edu.hk}
\affiliation{%
  \institution{The Hong Kong Polytechnic University}
  \city{Hong Kong}
  \country{China}
}

\author{Zhi Zhang}
\email{zhi271.zhang@connect.polyu.hk}
\affiliation{%
  \institution{The Hong Kong Polytechnic University}
  \city{Hong Kong}
  \country{China}
}

\author{Gong Chen}
\email{heinz@clozzz.com}
\affiliation{%
  \institution{FireTorch Partners}
  \city{Hong Kong}
  \country{China}
}

\author{Bruce X.B. Yu}
\email{xinboyu@intl.zju.edu.cn}
\affiliation{%
  \institution{ZJU-UIUC Institute, Zhejiang University}
  \state{Zhejiang}
  \country{China}
}

\author{Jiannong Cao}
\email{csjcao@comp.polyu.edu.hk}
\affiliation{%
  \institution{The Hong Kong Polytechnic University}
  \city{Hong Kong}
  \country{China}
}

\renewcommand{\shortauthors}{Trovato et al.}


\begin{abstract}
In the era of human–AI co-creation, the maxim ``knowing is easy, doing is hard'' is redefined. AI has the potential to ease execution, yet the essence of ``hard'' lies in who governs the translation from knowing to doing. Mainstream tools often centralize interpretive authority and homogenize expression, suppressing marginal voices. To address these challenges, we introduce the first systematic framework for redistributing authority in the knowing--doing cycle, built on three principles, namely contestability, agency, and plurality. Through interactive studies with 180 music practitioners, complemented by in-depth interviews, we demonstrate that these principles reshape human--AI authority relations and reactivate human creative expression. The findings establish a new paradigm for critical computing and human--AI co-creation that advances from critique to practice.
\end{abstract}



\maketitle

\section{Introduction}
\begin{CJK}{UTF8}{gbsn}
The maxim ``knowing is easy, doing is hard'' (\textit{知易行难}) has deep roots across philosophical traditions. Both the \textit{Book of Documents} (\textit{Shangshu})—with its maxim ``Knowing is not that difficult, but doing is'' (非知之艰，行之惟艰)—and Aristotle’s \textit{Nicomachean Ethics} underscore the gap between knowing (\textit{episteme}) and doing (\textit{praxis}), emphasizing that understanding without \textit{phronesis} (practical wisdom) cannot ensure effective action \cite{crisp2014aristotle}.
\end{CJK}

In today’s landscape, where AI is deeply embedded in human–AI co-creation, the gap between knowing and doing takes on a new form. We define \textit{knowing} as the capacity to access and represent knowledge with precision, and \textit{doing} as the situated translation of such knowledge into creative artifacts. Large language models (LLMs) demonstrate remarkable power in supporting knowing across domains \cite{epperson2025interactive, zheng2025customizing, shao2025unlocking, siddiqui2025script, zhong2024ai}. However, they provide limited support for doing; the dominant prompt--output paradigm of AI often deepens rather than bridges the gap that humans face in transforming knowledge into creative outcomes \cite{gpt4o, midjourney2024, suno2024}. For example, users are compelled to compress layered intentions into short prompts, erasing much of their tacit richness. Models, in turn, resolve ambiguity by defaulting to common training patterns, so doing becomes absorbed into system logic and users must adapt to it.

Within the creative domain of music \cite{krol2025exploring,cavez2025euterpen,lee2025mvprompt,choi2025understanding}, this gap is especially pronounced. Human composers often begin with layered and ambiguous intentions, such as evoking a sense of ``flowing melancholy'', which require iterative probing of notes, harmonies, and instrumentation to gradually transform into concrete artifacts. In contrast, when working with AI systems, these intentions must be compressed into precise and complete instructions, erasing much of their tacit richness because models cannot sense what ``flowing melancholy'' means. When faced with such underspecified requests, models routinely fall back on the most common patterns in their training data, defaulting to familiar keys, regular meters, and mainstream genres. As a result, the human process of doing becomes absorbed into system logic, reinforcing rather than resolving the knowing–doing gap.

This gap connects to two defining features of artistic creation. First, art admits no absolute ``right'' or ``wrong'' \cite{wittgenstein1966lectures}. Schoenberg’s twelve-tone system, once condemned for breaking tonal rules, later reshaped modern music \cite{hyde1985musical,marcus2016schoenberg}. Such vitality from ``rule-breaking'' shows that doing requires navigating ambiguity in ways that cannot be reduced to rule-based knowing. Second, artistic creation advances by ``leading the trend'' rather than ``obeying the majority'' \cite{bourdieu2018distinction}. When the Beatles incorporated classical strings into rock rhythms, they initially faced resistance, yet their experiments redefined the soundscape of popular music \cite{spitz2012beatles, everett1999beatles}. This illustrates that doing depends on exploration beyond existing patterns, whereas models trained on past data tend to reinforce them.

These two features reveal that the knowing–doing gap in the age of AI is not only a technical matter of limited support for creative practice but a structural matter of authority. When layered intentions are compressed into prompts and underspecified requests are resolved by defaulting to familiar keys, regular meters, or mainstream genres, it is the model rather than the human that decides how knowing is translated into doing. From this perspective, the dominant prompt--output paradigm of AI in human–AI co-creation produces three interrelated structural challenges. The first challenge is the interpretive monopoly in which the reasoning about how intentions are translated into outputs remains hidden inside the model and users have little ground to contest its decisions. The second challenge is the loss of agency in which doing is absorbed into defaults and users are limited to rephrasing prompts rather than shaping the generative process. The third challenge is homogenization in which marginalized expressions are normalized out as models regress to dominant conventions. These challenges recast the maxim ``knowing is easy, doing is hard'' as a struggle over authority, asking who has the authority to decide how knowing becomes doing.

To address these challenges of authority imbalance, we pose three core research questions.
\begin{itemize}[left=0pt]
    \item What conditions of contestability are needed to break AI’s interpretive monopoly and allow users to dispute the system’s logic of translating intentions into outputs?
    \item How can agency be restored so that users regain active control over how knowing is translated into doing?
    \item In what ways can plurality safeguard non-mainstream creative pathways and protect marginalized voices from systemic homogenization?
\end{itemize}

To examine these questions, we designed a systematic framework to operationalize and evaluate the three anti-oppressive principles. We recruited 180 music practitioners, including novices, intermediates, and advanced creators, for a six-week empirical study. The findings show that each principle directly addressed its guiding question. Contestability enabled users, particularly beginners, to recognize and dispute the system’s translation logic. Agency allowed users to shift from passively receiving results to actively shaping generative processes. Plurality provided protection for marginalized voices against homogenization. Together, these results demonstrate that anti-oppressive principles can effectively reconfigure human–AI power relations, showing that authority in the knowing–doing cycle can be redistributed through design.

Our paper advances three critical contributions to critical computing and creativity-justice–oriented HCI:
\begin{itemize}[left=0pt]
    \item We reframe human–AI co-creation by positioning interpretive authority as the central axis of the knowing–doing cycle, revealing how systems consolidate power through hidden interpretation and why authority must be treated as a matter of justice in creative technologies.
    \item We introduce contestability, agency, and plurality as anti-oppressive design principles that counter interpretive monopoly, default-driven normalization, and homogenization, protecting marginalized expressions while supporting equitable translation of abstract intentions into concrete doing.
    
    \item Through a study with 180 practitioners, we show that authority in the knowing–doing cycle is redistributable. Beginners learned to contest system-defined correctness, intermediates stabilized non-mainstream practices, and advanced practitioners defended minority aesthetics, reframing ``knowing is easy, doing is hard'' as a struggle over authority and a site where creativity justice becomes tangible.
\end{itemize}

\section{Background}
We situate our work at the intersection of two lines of research. The first examines how HCI systems distribute interpretive authority, showing that transparency alone does not redress structural asymmetries and that contestability and plurality are required. The second examines paradigms of human--AI co-creation, where interaction remains confined to outputs rather than reasoning, leaving users little influence over interpretive decisions. Taken together, these perspectives establish both the theoretical foundation and the application context for our study.

\subsection{Interpretive Authority in HCI}

HCI has long examined how sociotechnical systems distribute interpretive authority. Feminist and critical HCI show that design practices encode values and authority relations, often marginalizing users' voices or subordinating them to system logics \cite{bardzell2010feminist,sengers2005reflective,suchman2007human}. Work on algorithmic decision-making further demonstrates how opaque infrastructures constrain contestability and reproduce asymmetries among designers, systems, and end users \cite{eslami2016first,lyons2021conceptualising}. In response, scholars have proposed designs that redistribute authority by creating sites where users can question, intervene in, and co-define how knowledge is operationalized \cite{ehsan2021expanding,lee2019webuildai,green2019disparate,costanza2020design}.

Building on this foundation, recent HCI and critical computing emphasize that AI systems are infrastructures in which interpretive authority is actively designed and allocated. Ethics and design choices embed situated authority relations that shape whose voices are amplified and whose are marginalized \cite{tanksley2025ethics}. In creative domains, this manifests as risks of re-centering Western aesthetics and extracting value from creative labor while eroding authorship, alongside tensions over who defines system boundaries and responsibilities \cite{qadri2025ai,porquet2025copying,kollig2025fictional}. Beyond creativity, fairness-and-contestability research indicates that transparency primarily improves informational fairness, whereas structured avenues for contestation are required to enhance procedural fairness and perceived legitimacy \cite{vaccaro2019contestability,yurrita2023disentangling}.

A growing body of work makes AI reasoning more visible. Algorithmic advances include linear chain-of-thought prompting \cite{wei2022chain,wang2023selfconsistency,zhou2023leasttomost,zhang2024improvingcot}, structured search such as tree- and graph-based reasoning \cite{yao2023treeofthoughts,besta2024graphofthoughts}, and agent frameworks \cite{yao2023react,shinn2023reflexion}. HCI complements these with tools for inspection and comparison \cite{arawjo2024chainforge,kahng2024llmcomparator}, sensemaking interfaces \cite{gero2024sensemaking}, and verification-oriented pipelines \cite{zhao2023verifyandedit,jacovi2024reveal,xu2024symbcot}, as well as recent explorations of causal-pathway views, prompt critique, intent-aligned workflows, and interactive reasoning trees \cite{Zhong2024,KimTaeSoo2024,Zeng2024,Pang2025}. Taken together, these efforts establish a foundation for interactive representations of reasoning. However, visibility primarily supports informational understanding rather than procedural recourse. Without mechanisms that let users intervene in, revise, and steer the reasoning process, interface transparency risks leaving interpretive authority intact. Our work builds on this foundation by turning model reasoning into an actionable medium that users can inspect, edit, and contest, thereby redistributing interpretive authority in situated creative tasks.

At the infrastructure level, analyses of data practices and platform services show how upstream assumptions and one-size-fits-all architectures prefigure downstream defaults \cite{kapania2023hunt,lewicki2023out}. At the level of language and culture, studies document constraints arising from masculine defaults in datasets, gendered query formulations, and unmet needs around code-mixing; these findings evidence structural limits on expression, though linking them directly to uneven distributions of interpretive authority warrants further evidence \cite{kopeinik2023show,seaborn2023transcending,choi2023toward}.

Taken together, this body of work establishes that interpretive authority in AI-mediated creativity is not neutral and cannot be addressed by transparency alone. What is required are mechanisms for contestability, agency, and safeguards for plurality. Our study advances this agenda by showing that turning model reasoning into an actionable medium—one that users can inspect, edit, and contest—enables the redistribution of interpretive authority and resists the absorption of diverse practices into dominant defaults.

\subsection{Paradigms of Human–AI Co-Creation}
Research on human--AI co-creation has advanced through several paradigms, yet most retain the same structural orientation. Interaction occurs at the level of outputs rather than at the level of reasoning. Prompt-based systems allow users to provide natural language or visual descriptions, with the model returning a finished artifact \cite{mahdavi2024ai,Wang2024,Rajcic2024,Liu2023,GoloujehEA2024,liu2022design}. While these systems are flexible and accessible, they leave users navigating opaque prompt journeys with little ability to question how their intentions were interpreted. Steerable control interfaces introduce sliders, handles, or latent directions to adjust features such as style, pose, or tempo \cite{IntentTuner,DirectGPT,AdaptiveSliders,SketchFlex,FusAIn,Expandora,dang2022ganslider,evirgen2023ganravel,pan2023drag,shi2024dragdiffusion}. These interfaces provide direct manipulation but typically restrict users to surface-level adjustments while the interpretive logics of the model remain hidden. Iterative co-creation systems frame AI as a turn-taking partner in writing, drawing, and music \cite{Jakesch2023,Reza2024,Dhillon2024,Lin2024,He2025,Kim2025,Khan2025,Wang2025,zhang2022storydrawer,davis2025ai,wan2024felt,malandro2024composer}. Although this paradigm emphasizes collaboration, the exchange is still confined to outputs such as lines of text, brushstrokes, or notes, rather than the processes that produced them.  

Across these approaches, the knowing--doing cycle remains concealed. Users may rewrite prompts, adjust parameters, or respond to outputs, but they cannot directly access, contest, or redirect the model’s interpretive decisions. This asymmetry matters especially in creative domains where value lies not in correctness but in plurality, negotiation, and the ongoing transformation of ideas into artifacts. The gap in existing work is therefore not only technical but structural: co-creation is framed as interaction with outputs rather than as participation in the reasoning that shapes them. Our study addresses this gap by empirically testing whether and how the knowing--doing cycle can be made contestable, open to user agency, and protective of plurality.

\section{Method}
\subsection{Research Goal}
Our method is designed to operationalize the three research questions introduced in the introduction, examining whether the knowing--doing cycle in human--AI creative practice can be turned into a site of contestation over interpretive authority. While the maxim ``knowing is easy, doing is hard'' frames the long-standing gap of moving from abstract intention to concrete expression, the prompt--output paradigm transforms this gap into a structural imbalance where models monopolize interpretation, users adapt to defaults, and diverse expressions are normalized out. 

We therefore focus on three dimensions:  

\paragraph{Contestability.} Whether participants can see and dispute how their intentions are interpreted in the cycle, and whether such disputes alter subsequent outputs.  

\paragraph{Agency.} Whether participants can redirect the generative trajectory beyond prompt retries, and whether this redirection propagates faithfully through the knowing--doing cycle.  

\paragraph{Plurality.} Whether collaboration can sustain non-mainstream paths in the cycle, resisting regression to defaults and preserving diversity across iterations.

Together, these dimensions allow us to investigate whether the knowing--doing passage can shift from a system-controlled pipeline into a negotiated medium of co-creation, where interpretive authority is shared rather than monopolized.

\subsection{Comparable Conditions}

To probe contestability, agency, and plurality, we compared three system conditions that represent different ways of structuring the knowing--doing passage. All conditions were implemented on the same model backbone, GPT-4o \cite{gpt4o}, to avoid confounding by model capability. The difference lies only in how reasoning is surfaced and acted upon.

\paragraph{Baseline (prompt--output).} Participants typed prompts and received only final musical outputs. Reasoning was hidden and could not be inspected or contested. The only available form of control was to rephrase the prompt. This represents the dominant paradigm where knowing is reduced to prompt tokens and doing collapses into the system’s defaults.

\paragraph{Transparency-only.} Participants were shown a representation of the model’s intermediate reasoning. This reasoning surface could be reviewed but not edited. Users could observe how intentions were interpreted but had no means to intervene. This condition isolates the effect of visibility without agency, making contestability possible in principle but leaving agency absent.

\paragraph{Mutelier.} Music Atelier (Mutelier) served as an intervention to probe structural asymmetries. It externalized reasoning as an editable surface, enabled users to redirect the generative process, and issued drift notices with plurality-preserving branches when regression to defaults was detected. These affordances correspond to our three research questions: reasoning surface for contestability, process redirection for agency, and plurality safeguards for plurality. Unlike contestability and agency, which are properties of reasoning visibility and editability, plurality is defined at the level of outcomes. It concerns whether non-mainstream expressions are actively protected from collapsing into defaults. Only Mutelier introduced explicit safeguards for plurality, while the other two conditions left diversity vulnerable to normalization.

\begin{figure*}[ht]
    \centering
    \includegraphics[width=\textwidth]{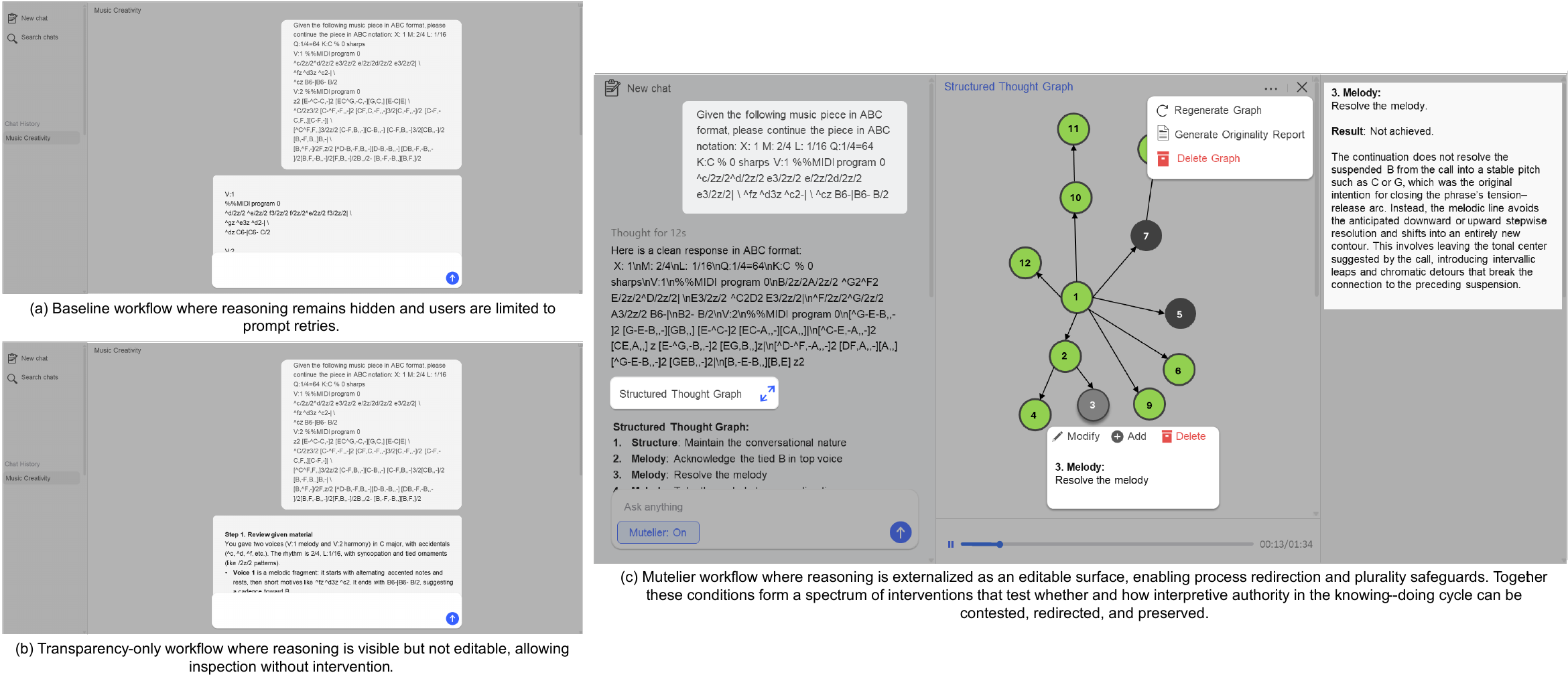}
    \caption{Comparable conditions. (a) Baseline prompt--output hides reasoning and limits users to prompt retries. (b) Transparency-only makes reasoning visible but not editable. (c) Mutelier externalizes reasoning as an editable surface, allows redirection, and provides plurality safeguards.}
    \Description{Comparable conditions. (a) Baseline prompt--output hides reasoning and limits users to prompt retries. (b) Transparency-only makes reasoning visible but not editable. (c) Mutelier externalizes reasoning as an editable surface, allows redirection, and provides plurality safeguards.}
    \label{fig:conditions}
\end{figure*}

Figure~\ref{fig:conditions} illustrates the three conditions. The baseline workflow (a) hides reasoning entirely, limiting users to prompt retries. The transparency-only workflow (b) makes reasoning visible but not editable, offering inspection without intervention. The Mutelier workflow (c) externalizes reasoning as an editable surface, supports redirection, and provides plurality safeguards. Together, these conditions establish a spectrum of interventions that allow us to test whether and how interpretive authority in the knowing--doing cycle can be contested, redirected, and preserved.

\subsection{Study Procedure}
\begin{figure*}[ht]
    \centering
    \includegraphics[width=\textwidth]{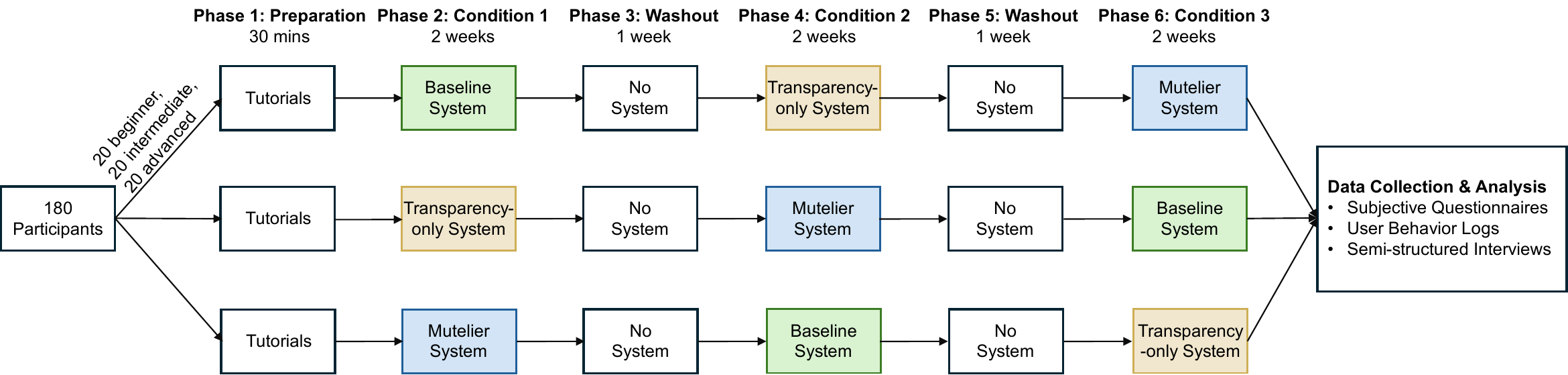}
    \caption{Study procedure with three conditions in counterbalanced order. Each condition lasted two weeks, followed by a one–week washout. Tasks included three briefs and one self–directed project. Data sources included outputs, interaction logs, questionnaires, and interviews.}
    \Description{Study procedure with three conditions in counterbalanced order. Each condition lasted two weeks, followed by a one–week washout. Tasks included three briefs and one self–directed project. Data sources included outputs, interaction logs, questionnaires, and interviews.}
    \label{fig:procedure}
\end{figure*}

We adopted a three–condition crossover design to investigate how contestability, agency, and plurality manifest across different interaction configurations. Each participant experienced Baseline, Transparency–only, and Mutelier in counterbalanced order. All conditions shared the same model backbone so that only interaction affordances varied. Each condition lasted two weeks and was followed by a one–week washout period in which participants returned to their normal creative routines. This design reduced carryover effects and allowed us to compare conditions within the same participant. The overall sequence is summarized in Figure~\ref{fig:procedure}.

At the beginning of each condition, participants were given three creative briefs designed by the research team (melody–to–accompaniment generation, continuation of a given excerpt, and text–to–music composition) as well as one self–directed project drawn from their own practice. The self–directed task ensured that participants could situate the system in ongoing creative work rather than only responding to predefined prompts. Table~\ref{tab:procedure} details the phases, durations, and data collected across the study.

During each condition, participants were free to use the assigned system iteratively in their creative process. They could generate outputs, inspect reasoning surfaces when available, and, depending on the condition, contest or redirect the system’s reasoning. All user interactions were automatically logged, including generated outputs, reasoning surfaces, edit events, and branching choices. These logs provided a trace of how participants negotiated or resisted the system’s defaults over time.

At the end of each condition, participants completed a 15–item questionnaire on transparency, controllability, and collaborative fairness. In addition, a subset of participants (two from each expertise group, six in total) took part in in–depth semi–structured interviews. These interviews probed how participants articulated tacit intentions, perceived defaults, and exercised contestation in practice. Interview data were triangulated with interaction logs and survey results to provide a multi–layered picture of how the knowing--doing cycle was reshaped across conditions.

\begin{table*}[h]
\centering
\caption{Study phases, duration, main activities, and data collected under the three–condition crossover design.}
\label{tab:procedure}
\begin{tabular}{cp{3cm} p{1cm} p{4cm} p{4cm}}
\toprule
\textbf{Index} & \textbf{Phase} & \textbf{Duration} & \textbf{Main Activities} & \textbf{Data Collected} \\
\midrule
1 & Preparation & 30 mins & Consent, demographics, tutorials & Demographic data; baseline survey \\
2 & Condition~1 & 2 weeks & Free use of first system (Baseline / Transparency-only / Mutelier) for briefs and one self-directed project & Interaction logs; questionnaire; 2 interviews/group \\
3 & Washout & 1 week & No study systems; participants continue normal music-making & Reflections on practice during washout \\
4 & Condition~2 & 2 weeks & Switch to second system; repeat tasks & Interaction logs; questionnaire; 2 interviews/group \\
5 & Washout & 1 week & Same as Washout~1 & Reflections on practice during washout \\
6 & Condition~3 & 2 weeks & Switch to final system; repeat tasks & Interaction logs; questionnaire; 2 interviews/group \\
\bottomrule
\end{tabular}
\end{table*}

\subsection{Participants}

We recruited 180 participants in collaboration with a music–technology company that provides tools and educational resources for composers, producers, and educators. Recruitment followed three channels: conservatories, online and offline musician communities, and professional networks of freelance composers and educators. Eligibility required active music–making within the past month, verified through course enrollment, ongoing projects, or publicly available works. 

Participants were stratified into three groups by years of verified compositional practice. This stratification was not only demographic but also critical for examining how structural imbalances in the prompt--output paradigm shape creators’ positions of authority.  

\paragraph{Beginners (0–1 year).} Typically students or hobbyists with limited training, this group represents those most vulnerable to interpretive monopoly. In current systems, their layered intentions are easily compressed into oversimplified prompts, leaving little room to dispute or redirect. We therefore focused on whether the intervention enabled them to contest the system’s reasoning and reclaim agency.  

\paragraph{Intermediates (1–10 years).} These participants could complete original works and were proficient with digital audio workstations. Situated between novices and professionals, they occupy a transitional position where creative intentions are more developed but still frequently subordinated to system defaults. Their engagement allowed us to observe how agency is consolidated or undermined in mid-level practice.  

\paragraph{Advanced (>10 years).} Professional musicians, freelance composers, and educators with established portfolios. This group often resists homogenization and seeks to preserve minority aesthetics such as irregular meters or experimental harmonies. For them, plurality safeguards are especially relevant, as they reveal whether the intervention can protect non-mainstream expressions against regression to defaults.  

By examining these three groups together, we could test whether the intervention altered structural inequalities in differentiated ways: whether it gave novices the ability to contest, enabled intermediates to redirect, and supported advanced users in preserving plurality. Table~\ref{tab:participants} summarizes the demographic distributions.

\begin{table*}[h]
\centering
\caption{Participant demographics by experience group. Values show $n$ (male/female), age range with mean, years of experience (mean$\pm$SD), primary roles, and AI familiarity.}
\label{tab:participants}
\begin{tabular}{lccccp{2.8cm}p{2.4cm}}
\toprule
\textbf{Group} & \textbf{$n$ (M/F)} & \textbf{Age Range} & \textbf{Mean Age} & \textbf{Years Exp.} & \textbf{Primary Role} & \textbf{AI Familiarity} \\
\midrule
Beginner 
& 60 (29/31) 
& 19--23 
& 20.2 
& $0.6 \pm 0.3$ 
& Student 88\%, Hobbyist Composer 12\% 
& Frequent $\geq$2/wk: 34\%; Occasional: 66\% \\

Intermediate 
& 60 (30/30) 
& 22--35
& 25.4 
& $5.1 \pm 0.9$ 
& Freelance Composer 42\%, Content Creator (Music) 34\%, Educator 18\%, Others 6\%
& Frequent: 82\%; Occasional: 18\% \\

Advanced 
& 60 (32/28) 
& 30--55 
& 36.1 
& $13.6  \pm 2.7$ 
& Musician 34\%, Freelance Composer 32\%, Educator 26\%, Others 8\% 
& Frequent: 88\%; Occasional: 12\% \\
\midrule
\textbf{Total} 
& \textbf{180 (91/89)} 
& 19--55 
& 27.2 
& --- 
& --- 
& --- \\
\bottomrule
\end{tabular}
\end{table*}

\subsubsection{Interview Sample}
To deepen the analysis beyond survey scores and system logs, we conducted follow–up interviews with six participants drawn from different strata of experience. These cases were selected for their illustrative value in showing how contestability, agency, and plurality were enacted from distinct positions of authority. Table~\ref{tab:interview_sample} summarizes their backgrounds.

\begin{table*}[h]
\centering
\caption{Profiles of six participants selected for in–depth interviews.}
\label{tab:interview_sample}
\resizebox{\textwidth}{!}{%
\begin{tabular}{lcccp{3.2cm}p{2.8cm}}
\toprule
\textbf{ID} & \textbf{Group} & \textbf{Years Exp.} & \textbf{Role} & \textbf{Genre / Style Focus} & \textbf{AI Familiarity} \\
\midrule
B1 & Beginner & 0.5 & Conservatory freshman (Composition) & Classical piano, early composition exercises & Occasional \\
B2 & Beginner & 1 & Computer–music hobbyist & Electronic sketches, DAW experimentation & Frequent \\
I1 & Intermediate & 3 & Freelance songwriter & Pop and crossover genres & Frequent \\
I2 & Intermediate & 5 & Middle–school music teacher & Pedagogical arrangements, choral works & Occasional \\
A1 & Advanced & 20+ & Local professional musician & Cantonese opera, regional folk traditions &  Frequent \\
A2 & Advanced & 30+ & Conservatory professor (Composition) & Contemporary classical, experimental harmony & Occasional \\
\bottomrule
\end{tabular}%
}
\end{table*}

\subsection{Measures and Data Sources}
To examine whether the intervention shifted structural imbalances in the prompt--output paradigm, we combined quantitative surveys, system logs, and qualitative interviews. Each source of data was mapped to the three research questions (RQ1 Contestability, RQ2 Agency, RQ3 Plurality), ensuring that the evidence reflected not efficiency or accuracy, but the redistribution of interpretive authority in the knowing--doing cycle.

\paragraph{Quantitative survey.} 
At the end of each condition, participants completed a 15–item Likert questionnaire (1 = strongly disagree, 7 = strongly agree). The items were grouped into three dimensions aligned with our research questions: Contestability (C1–C5), probing whether users could see, question, and influence the system’s reasoning; Agency (G1–G5), probing whether users could intervene directly in the knowing--doing process rather than adapt through prompt retries; and Plurality (P1–P5), probing whether non–mainstream ideas were protected against regression to defaults and whether multiple creative logics could coexist. The full mapping of dimensions, sub–aspects, and items is shown in Table~\ref{tab:survey_items}.

\paragraph{System logs.} 
During all tasks, the system recorded detailed interaction traces. These included the reasoning surfaces displayed, the edits users made (deletions, insertions, annotations), the disputes raised against model assumptions, and the adoption or rejection of plurality safeguards. Rather than analyzing logs for efficiency, we treated them as evidence of how contestability, agency, and plurality were enacted in practice. For example, whether disputes were propagated across iterations, whether redirections successfully reshaped outputs, and whether branches preserved minority aesthetics.

\paragraph{Interviews.} 
To complement the surveys and logs, we conducted in–depth interviews with a subsample of six participants (two from each experience group). These interviews probed how participants experienced authority asymmetries and how they perceived their role shifting when reasoning became contestable, processes became redirectable, and plurality safeguards were available. The interviews also allowed us to capture reflections that logs and surveys could not, such as the affective weight of having one’s intentions recognized or the frustration of being normalized out. 

Together, these data sources allowed us to triangulate across trends, behaviors, and lived experiences, grounding our critique not in abstract claims but in situated evidence of how interpretive authority was redistributed across the three conditions.

\begin{table*}[h]
\centering
\caption{Mapping of evaluation dimensions, sub-aspects, and questionnaire items for the three critical RQs. Each item was rated on a 7-point Likert scale (1 = strongly disagree, 7 = strongly agree).}
\label{tab:survey_items}
\begin{tabular}{p{3cm} p{3.6cm} p{0.9cm} p{5.5cm}}
\toprule
\textbf{Dimension} & \textbf{Sub-aspect} & \textbf{Code} & \textbf{Survey Item} \\
\midrule
\multirow{5}{*}{Contestability (RQ1)}
& Reasoning visibility & C1 & I could see how the system interpreted my intentions. \\
& Dispute entitlement & C2 & I felt I had the right to question the system’s interpretation. \\
& Dispute impact & C3 & My disputes visibly influenced subsequent reasoning or outputs. \\
& Targeted contestation & C4 & I could challenge specific steps of the system’s reasoning rather than rejecting the whole output. \\
& Contest legitimacy & C5 & Contesting the system felt like a legitimate part of the creative process, not an error. \\
\midrule
\multirow{5}{*}{Agency (RQ2)}
& Process steering & G1 & I felt able to steer how my ideas were turned into music. \\
& Intervention fidelity & G2 & My interventions were accurately carried into subsequent outputs. \\
& Control granularity & G3 & I could act on fine–grained aspects of the reasoning or music. \\
& Active intervention & G4 & I did not feel restricted to prompt retries but could actively intervene in the process. \\
& System responsiveness & G5 & The system responded in real time to my interventions. \\
\midrule
\multirow{5}{*}{Plurality (RQ3)}
& Protection of minority ideas & P1 & My unconventional or minority ideas were preserved rather than normalized out. \\
& Default resistance & P2 & The system helped me avoid unwanted regression to common defaults (e.g., C major, 4/4). \\
& Outcome diversity & P3 & The outputs supported multiple valid interpretations of my intent, not just one. \\
& Rule-breaking allowance & P4 & The system allowed space for rule-breaking or experimental expressions. \\
& Plural authorship recognition & P5 & I felt that diverse creative logics (mine and the system’s) could coexist in the final outcome. \\
\bottomrule
\end{tabular}
\end{table*}

\begin{figure*}[h]
    \centering
    \includegraphics[width=\linewidth]{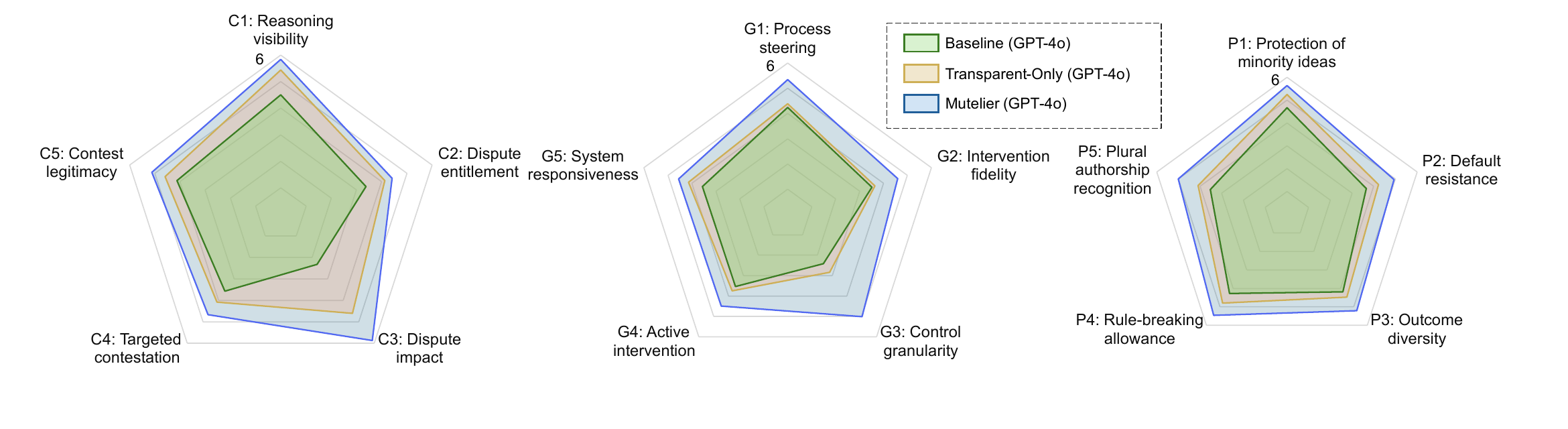}
    \caption{Radar plots of mean ratings for three evaluative dimensions aligned with the research questions: Contestability (RQ1), Agency (RQ2), and Plurality (RQ3). Results across Baseline, Transparency-only, and Mutelier show relative gains when reasoning is made contestable, though structural asymmetries are alleviated rather than resolved.}
    \Description{Radar plots of mean ratings for three evaluative dimensions aligned with the research questions: Contestability (RQ1), Agency (RQ2), and Plurality (RQ3). Results across Baseline, Transparency-only, and Mutelier show relative gains when reasoning is made contestable, though structural asymmetries are alleviated rather than resolved.}
    \label{fig:results_overview}
\end{figure*}

\begin{table*}[h]
\centering
\caption{Manifestations of structural imbalances across groups and differences in the practice of contestability, agency, and plurality.}
\label{tab:rq_comparison}
\begin{tabular}{p{2.1cm}p{3.5cm}p{3.5cm}p{3.5cm}}
\toprule
\textbf{Group} & \textbf{RQ1 Contestability} & \textbf{RQ2 Agency} & \textbf{RQ3 Plurality} \\
\midrule
\textbf{Beginners} &
Hidden reasoning forces them into ``guessing the right prompt''. Editability triggers an awareness of contestation—their first recognition that system assumptions can be challenged—yet uncertainty remains about when and how to intervene. &
Repetitive prompt trials strip agency; redirection provides ``entry-level empowerment'' but also introduces new cognitive load in deciding what to edit. &
Plurality is invisible under defaults; safeguards foster an awareness of difference as legitimate, but beginners hesitate to enact this recognition consistently. \\
\midrule
\textbf{Intermediates} &
They notice how unconventional ideas are disciplined into ``corrections''. Contestability becomes a practice of intent preservation, defending their direction through routinized action but requiring constant vigilance. &
They resist systemic drift through redirection, achieving process control, yet expend effort to withstand mainstream erosion. &
Homogenization is strongly felt; safeguards support the stabilization of non-mainstream directions, though at the risk of over-relying on the system’s definition of ``difference''. \\
\midrule
\textbf{Advanced} &
They anticipate overwriting but cannot resist it. Contestability becomes an act of authority defense, preventing system dilution of professional knowledge, and at the same time it can reinforce the expert–non-expert divide. &
Defaults undermine professional authority; redirection enables expertise reconstruction, though tensions with system arbitrariness persist. &
They focus on protecting minority aesthetics; safeguards function as a plurality shield against erasure, though cultural legitimacy remains contested. \\
\bottomrule
\end{tabular}
\end{table*}

\section{Results}
\subsection{Results Overview}
Across the three conditions, a consistent pattern emerges. Hidden reasoning leaves users subordinated to system defaults; visible but uneditable reasoning provides only partial relief, enabling diagnosis without intervention; only editable reasoning surfaces allow contestation, redirection, and the protection of minority aesthetics. This does not erase authority asymmetries but reconfigures them. Beginners recognize latent authority, intermediates consolidate direction while facing constant vigilance, and advanced users defend minority aesthetics while negotiating tensions with systemic defaults. Figure~\ref{fig:results_overview} presents quantitative trends across the three conditions, and Table~\ref{tab:rq_comparison} highlights group-specific manifestations of contestability, agency, and plurality.

\subsection{RQ1: Contestability—Challenging the monopoly of interpretation}
\subsubsection{Critical Hypothesis}  
Under the prompt--output paradigm, interpretive authority is held by the model. Users cannot see or dispute how intentions are translated into outputs, so they adapt to defaults rather than challenge them. Contestability is a matter of authority, not only feedback. If reasoning becomes visible and addressable, users can challenge the model’s framing and turn the knowing--doing passage from a closed pipeline into a site of shared authority.

\subsubsection{Evidence chain}
Survey data show a clear gradient. Contestability scores were highest under Mutelier, intermediate under Transparency-only, and lowest under Baseline (Fig.~\ref{fig:results_overview}, left panel). Visibility raised awareness of hidden assumptions, but only visibility with disputability changed who could act on them.

\paragraph{Beginners. Contesting the system’s authority to define gentle.}
For novices, contestability marked a first experience of authority. Participant B1, a first–year conservatory student, requested a gentle but slightly unstable piano phrase. Under the baseline system the output defaulted to a slow arpeggio in C major. She recalled
\begin{quote}
``I imagined something like waves, but what I got was just plain broken chords in C. It was so `correct' that it lost the shakiness I wanted. I thought maybe I didn’t explain well, so I tried typing `fragile' and then `wavering', but it still gave me the same safe chords. It felt like no matter what I said, the system decided for me what ‘gentle’ means.’’
\end{quote}

With Mutelier, she saw the reasoning surface display the link from gentle to slow tempo in C major to arpeggiated chords. The interaction logs captured the following trajectory:
\begin{itemize}
    \item Initial intent: ``gentle but slightly unstable, like breathing''  
    \item System interpretation: ``gentle = slow tempo; unstable = add suspension (resolved)''
    \item Dispute: ``Do not resolve the suspension; keep it hanging''
    \item Action: Deleted the ``resolution'' node, added note: ``leave second unresolved''  
\end{itemize}

The next version kept unresolved seconds rather than resolving to consonant chords. She said: 
 
\begin{quote}
``When I removed that ‘resolve suspension’ step, the sound finally felt closer to what I imagined—gentle but with a small tremble. It was the first time I felt the system didn’t overwrite my idea but let me insist on it.''
\end{quote}
The logs confirm that her dispute propagated and shaped the subsequent output.

\paragraph{Intermediates. Resisting a disciplinary reading of irregularity.}
For participants with several years of practice, contestability was enacted as a authority practice, both to resist the system’s quiet disciplining of irregular ideas and to refine their own intentions. Unlike beginners, who often discovered for the first time that they could object, intermediates used objections both to preserve non-standard techniques and to refine their own intentions.

Participant I1, with four years of songwriting experience, explained how rhythmic irregularity was repeatedly smoothed out:  
 
\begin{quote}
``I asked for a 7/8 groove because I wanted that uneven, dance-like push. The first version [Baseline] always flattened it back into 4/4, like the machine thought I was making a mistake. In the second version [Transparency-only], I could actually see the step saying ‘irregular rhythm → normalize to 4/4,’ and even some other things I didn’t want. But when I looked at all those lines, I didn’t know where to begin. It just drifted further and further away from my idea, and I had no way to stop it.''
\end{quote}

Participant I2 described contestability in more reflective terms:  
\begin{quote}
``I wanted to keep a harsh ninth chord, but the first version [Baseline] softened it automatically. In the editable one [Mutelier], I flagged that step, and then I realized maybe I hadn’t explained myself well. So I rewrote it as: ‘leave dissonance unresolved for tension.’ The system adjusted, but the important part was that I had to explain why the dissonance mattered. The dispute wasn’t only against the AI—it was also against my own vagueness.’’
\end{quote}

\paragraph{Advanced. Defending the legitimacy of minority traditions.}
For experienced practitioners, contestability was less about discovery or reflection and more about defending their interpretive authority. These users already anticipated that the system would overwrite unusual requests, yet in prior conditions they had no means to directly reject the model’s framing. With contestability, they could finally dispute and overturn specific interpretive steps that mischaracterized their intentions.

Participant A2, a professional musician with more than two decades practice, recalled:  
\begin{quote}
``I asked for a Cantonese opera style phrase with open fifths and sliding ornaments, because that timbre is central to the tradition. The baseline system kept replacing the slides with fixed pitches and filled the harmony into full triads. In another system [Transparency-only], I could see the step written out ‘open fifth → expand to triad’—but I could only watch it happen again and again, even when I explicitly typed ‘do not expand’. With the editable function, I deleted that step and added ‘retain open fifths, keep sliding ornaments’. The next version did not fully respect this choice, yet it allowed me to directly modify the reasoning itself. For me it was not about making the music sound nicer, but about insisting that the AI stop redefining what counts as legitimate in Cantonese opera.''
\end{quote}

\subsubsection{Critical Conclusion}  
Across all levels of expertise, the findings confirm that contestability transforms the knowing--doing passage from a closed pipeline into a site of negotiation. Beginners experienced a moment of authority awakening, discovering that they could object to the model’s hidden interpretations. Intermediates enacted contestability as an authority practice, using disputes to resist the quiet disciplining of irregular ideas while refining their own intentions. Advanced practitioners defended their authority, blocking the system from recasting their expertise as mistakes. Together, these trajectories show that contestability is not a minor usability enhancement but a structural reallocation of interpretive authority. By making reasoning visible and disputable, users reclaimed the right to challenge the model’s framing and to reshape the process on their own terms. The evidence demonstrates that the interpretive monopoly of the prompt--output paradigm is not inevitable but corrigible, and that contestability provides a concrete mechanism for redistributing authority in AI-mediated creativity.

\subsection{RQ2: Agency—Regaining control over the creative process}  
\subsubsection{Critical Hypothesis}
Under the prompt--output paradigm, users are stripped of agency. The passage from intention to outcome is taken over by system defaults, leaving users confined to endless prompt retries without any way to intervene in the process itself. Agency is therefore not a matter of convenience but of authority. It decides who has the right to shape the creative trajectory. If reasoning is not only visible but also editable, users can redirect the generative process, turning doing from a system-controlled pipeline into a negotiated practice. Restoring agency means redistributing the authority to steer creative outcomes.

\subsubsection{Evidence chain}

Survey results show a clear stratification of agency scores across the three conditions. Scores were lowest under Baseline, where participants overwhelmingly reported that the process was ``taken away'' by the system. Transparency-only improved slightly, since users could diagnose problems but remained unable to change them. Scores were highest under Mutelier, where participants reported actively redirecting the reasoning process, preventing outputs from sliding back into defaults. This trend is illustrated in the middle panel of Fig.~\ref{fig:results_overview}.

\paragraph{Beginners. From passive adaptation to taking control of the process.}
For novice participants, the restoration of agency marked an entry-level empowerment. Under Baseline they often felt helpless, confined to trying different prompts without effect.

B2, a computer music hobbyist, reflected on his experience with Mutelier:
\begin{quote}
``I wrote 'light and bouncy', but the system always gave me the same pop chord routine. I thought maybe I wasn’t clear, so I tried words like 'playful' or 'bouncy', but it still came out the same. I felt like I had no way to stop it from running back to its habits. Then I saw the step 'light to pop chord progression'. I deleted it and wrote 'add syncopation, no pop chords'. The next version finally followed me instead of dragging me along. For the first time I felt I was steering the process, not just accepting it.''
\end{quote}
This illustrates the shift from passive adaptation to active participation.

\paragraph{Intermediates. Resisting the quiet disciplining of the process.}
For mid-level creators, agency restoration was less about discovery and more about consolidating practice. They already had clear intentions, but under Baseline their work was often ``normalized'' back into mainstream conventions.

I1, a freelance composer, described how he used redirection:
\begin{quote}
``I deleted the step 'normalize to 4/4' and replaced it with 'keep 7/8 pulse'. The next version finally stayed in 7/8. To me it wasn’t just a tweak—it was pulling my practice back, stopping it from sliding into the mainstream.''
\end{quote}

\paragraph{Advanced. Defending the integrity of traditional practice.}
For experienced practitioners, agency restoration was less about learning to intervene and more about defending their authority. They often anticipated that the system would ``correct'' minority or experimental techniques, but under Baseline could only watch their practices be overwritten.

A1, a local musician who composes for government festivals, explained:
\begin{quote}
``At festivals we need the local flavor. I wrote melodies with our traditional sliding notes and used open fifths for that spacious sound. But the Baseline system kept fixing the slides into rigid pitches and filling the open fifths into full triads, as if I was making a mistake.''
\end{quote}
For A1, the editable process became a way to insist that local aesthetics were not ``errors'' but legitimate practices.

\subsubsection{Critical Conclusion}  
Across groups, restoring agency transformed doing from a one-way pipeline into a shared process. Beginners experienced agency recovery as an authority awakening, discovering they could alter the process rather than endlessly retry prompts. Intermediates enacted it as a authority practice, redirecting the system to consolidate their own intentions against normalization. Advanced practitioners defended their authority, blocking the system from rewriting their expertise as error. Together, these trajectories confirm that agency is not a minor usability feature but a structural site of authority: when reasoning is editable, users reclaim the right to steer the knowing--doing cycle, and the model ceases to hold unilateral control over how intentions become actions.

\subsection{RQ3: Plurality—Resisting the homogenization of expression}  
\subsubsection{Critical Hypothesis}  
Under the prompt--output paradigm, plurality is systematically undermined. Models are trained to privilege majority defaults and regress toward familiar conventions, collapsing diverse or minority aesthetics into homogenized forms. Plurality is not only a matter of stylistic variation but of legitimacy whose sounds and traditions are recognized as valid. If explicit safeguards are introduced, systems should resist this collapse, protecting non-mainstream paths and allowing users to sustain diverse forms of expression.

\subsubsection{Evidence chain}
Survey data confirmed that plurality protection was weakest under Baseline, slightly improved under Transparency-only, and strongest under Mutelier (Fig.~\ref{fig:results_overview}, right panel). Visibility of reasoning helped participants recognize when non-mainstream features were erased, but only plurality safeguards prevented this erasure from dominating the final outputs.

\paragraph{Beginners. Contesting the myth of a single ``correct'' option.} Newcomers often interpreted defaults such as piano timbre or C major as the only valid choice. Several even self-corrected their intentions to match what the system produced, gradually accepting homogenized results as ``proper music''. Participant B2, a computer music hobbyist, reflected:  
\begin{quote}
``I asked for something bright with unusual colors, but it always gave me piano in C. After a few tries, I thought maybe that’s what `bright' means, so I stopped asking for different instruments. It felt like the system was teaching me the right answer''.
\end{quote}

\paragraph{Intermediates. Resisting gradual domestication of irregularity.} For participants with several years of practice, the issue was not immediate collapse but incremental normalization. Experimental timbres or irregular forms were gradually pulled back into mainstream conventions across iterations. Participant I1, a freelance creator, described this drift:  
\begin{quote}
``At first it kept my strange synth texture, but each new version smoothed it more, until it just sounded like regular pop pads. I could sense it was slipping away, but I couldn’t stop it. It was like being tamed bit by bit.''
\end{quote}
With safeguards, he could branch back toward his original intent, preventing the iterative erosion of plurality.

\paragraph{Advanced. Defending minority traditions from erasure.} Experienced practitioners highlighted how minority or traditional aesthetics were directly rejected as ``errors'' or ``unrecognized forms''. Participant A2, a local musician who composes for government celebrations, recalled:  
\begin{quote}
``the AI system was about erasing the legitimacy of our culture''.
\end{quote}
Plurality safeguards allowed him to explicitly preserve open fifths and slides, forcing the system to treat them not as mistakes but as valid directions.

\subsubsection{Critical Conclusion}  
The findings show that homogenization is not a side effect but a structural feature of the prompt--output paradigm. Beginners internalized defaults as correctness, intermediates struggled against gradual domestication, and advanced practitioners confronted outright delegitimation of minority aesthetics. Plurality safeguards did not guarantee perfect fidelity, but they established space where non-mainstream expressions could persist rather than collapse into the majority. The evidence demonstrates that protecting plurality is essential for redistributing legitimacy in AI-mediated creativity, ensuring that diverse artistic practices are not overwritten by dominant norms.

\section{Discussion}
Our findings go beyond improvements in usability metrics to expose the structural limitations of current paradigms for human--AI collaboration. The dominant prompt--output model privileges efficiency and rule-following, yet structurally reproduces a divide between knowing and doing, compressing complex intentions and silencing nuanced expression. Through the design and validation of \textit{Mutelier}, we argue that a shared knowing--doing medium is not merely a design enhancement but a critical intervention. It redistributes interpretive authority, restores space for negotiation, and challenges the rationalist framing of creativity as singular and closed.

\subsection{Reconstructing the Knowing--Doing Medium as Critical Intervention}

Our study treats the knowing--doing passage as a site of authority rather than a neutral pipeline. The prompt--output paradigm centralizes interpretive authority in the model and normalizes outcomes toward majority defaults. The intervention introduced in this paper functions as an empirical probe that makes the passage visible, editable, and safeguarded, which allows us to observe how authority can shift when users dispute, redirect, and preserve alternative logics.

\paragraph{Contestation redistributes interpretive authority.} Across conditions, visibility without addressability raised awareness but did not change who could act. When reasoning became a surface that users could contest, disputes were no longer complaints at the margins but operations that altered subsequent steps. Beginners experienced a first awakening of authority when they could refuse the model’s definition of their intentions. Intermediates practiced authority by challenging disciplinary framings such as the automatic smoothing of irregular rhythm and unresolved dissonance. Advanced practitioners used contestation to block the recoding of domain expertise and minority traditions as error. These trajectories show that contestability is not a convenience feature. It is a mechanism that moves interpretive authority from a model monopoly to a negotiated relation.

\paragraph{Process redirection restores agency in doing.} Under the prompt--output paradigm, doing is absorbed into defaults and the user is confined to prompt retries. Logs and interviews show that editability of reasoning steps changes the locus of control. Beginners moved from guessing words to steering the path of action. Intermediates consolidated practice by holding ideas on course against gradual drift toward mainstream forms. Advanced users defended the integrity of established methods when the system attempted to overrule them. Agency therefore becomes a property of process control rather than an attribute of individual skill, and it can be designed for.

\paragraph{Plurality requires explicit protection to resist homogenization.} Plurality is not guaranteed by visibility or even by contestation alone. In our data, beginners often internalized defaults such as piano in C major as correctness. Intermediates encountered incremental domestication where experimental timbres and structures were smoothed across iterations. Advanced practitioners faced delegitimation when minority aesthetics were rewritten as unrecognized or incorrect. Safeguards that expose drift and offer branches made room for non-mainstream paths to persist. The lesson is that plurality is a question of legitimacy. Systems must recognize and protect alternative logics rather than treat them as deviations to be corrected.

\paragraph{Implications.} The findings suggest three principles. Make knowing contestable by rendering reasoning into an inspectable medium that supports objection and evidence. Restore agency by enabling edits that redirect the process so users can shape how intentions become actions. Protect plurality by detecting regressions to defaults and scaffolding branches that sustain minority aesthetics. Together these principles treat the knowing--doing passage as a negotiated medium where authority is shared. Evaluating authority redistribution requires measures beyond accuracy and preference. Our approach combined questionnaire trends with interaction traces of disputes, edits, and plurality branches, and with interviews that located these actions within participants’ positions in practice. This combination surfaces how different users meet the same system from unequal starting points and how design can redress or reproduce those inequalities.

\subsection{Challenging Rationalist Bias and Redistributing Creative Authority}
A distinctive feature of artistic creation lies in its conscious transcendence of rational logic. This is not a rejection of reason itself, but a critique of the rationalist bias that reduces creativity to optimization and coherence. Rationality provides necessary structure---from Kant’s account of aesthetic judgment as \textit{purposiveness without purpose} to Dewey’s \textit{view of art} as a heightened, integrated form of experience---yet art often achieves its effect precisely by departing from logic \cite{kant2000critique,dewey2024art}. For example, a melancholic melody may contain fleeting passages of brightness that intensify the sadness; a motif may gain authority through repetitive accumulation; seemingly incompatible tonalities may be juxtaposed to generate resonance. From the standpoint of rational optimization, such strategies appear ``inefficient'' or ``illogical'', yet they are the core of artistic expression---a point stressed from Schoenberg’s aesthetics to Meyer’s theory \cite{schoenberg1984style,meyer1954emotion}. The vitality of art emerges through its surpassing of rationalist constraints.

Current AI systems inherit and reinforce this rationalist bias. They privilege defaults, optimize away irregularities, and normalize outputs toward majority conventions. What emerges is not simply a technical limitation, but a political question: who has the authority to define whether dissonance is an ``error'' or a deliberate expression, whether repetition is ``derivative'' or a strategy of accumulation? Following Foucault’s notion of \textit{power/knowledge} and Ranci\`ere’s account of the \textit{distribution of the sensible}, AI systems do not neutrally translate knowing into doing; they configure which voices and which aesthetics are legitimized, and which are marginalized \cite{foucault2020power,ranciere2013politics}.  

Our study shows that these asymmetries can be made visible and contested. Beginners, often the most vulnerable, discovered they could dispute the system’s hidden assumptions rather than adapt to them. Intermediates used redirection to consolidate their practice and resist the quiet drift toward mainstream norms. Advanced practitioners defended minority traditions against being overwritten as ``errors''. Across these trajectories, contestability, agency, and plurality emerge not as usability features but as political dimensions of design.  

The broader significance is that the knowing--doing passage itself must be understood as a contested medium. When reasoning is exposed, editable, and safeguarded against homogenization, creative authority is redistributed: users reclaim the right to contest interpretations, to redirect processes, and to preserve non-mainstream expressions. In this sense, the critical task of human--AI collaboration is not to perfect rational optimization, but to resist rationalist bias and to reconstruct creation as a negotiated practice where authority is shared rather than imposed.

\subsection{Ethics Considerations}
This study involved 180 music practitioners across different levels of expertise in a six–week user study. Local regulations do not require formal ethics review for this type of study. Nevertheless, all participants were informed of the study aims, procedures, and their rights prior to participation, and written consent was obtained. Participation was voluntary, with the option to withdraw at any time without penalty. 

We took measures to minimize risk by ensuring that tasks were aligned with participants’ normal creative practices and did not involve sensitive personal information. Interaction logs and survey responses were anonymized and stored on secure servers accessible only to the research team. In reporting results, no identifiable information has been disclosed, and illustrative quotes from interviews were pseudonymized. 

Beyond human–subjects protection, we recognize broader ethical considerations in the domain of creativity-support systems. Our design foregrounds the redistribution of interpretive authority to counteract potential marginalization of users’ voices, aligning with ACM’s general principles of fairness, accountability, and respect for diverse practices. This ethical orientation complements our technical contributions by highlighting that interventions in human–AI co-creation are not value–neutral but actively shape the conditions of creative legitimacy.


\subsection{Limitations and Future Work}
While our study spans participants across different levels of compositional maturity, it remains situated within the domain of music. This limitation is not only methodological but structural. Different creative fields (e.g., visual design, creative writing, painting) encode distinct knowing--doing tensions and enact distinct distributions of authority. A ``color--composition'' graph in painting, for instance, may surface disputes over perspective and symbolism, which are qualitatively different from the harmony disputes we observed in music. Future work should therefore investigate not whether our intervention generalizes as a tool, but whether it can continue to reveal and redistribute interpretive authority in domains with different cultural and aesthetic logics.

The relatively lower ratings from advanced users also raise critical questions. Rather than interpreting this as a design flaw, it reflects how existing paradigms already privilege expert intuition while disproportionately marginalizing novices. Our intervention produced the largest redistributive effects for those with the least authority, yet future research must also consider how to support expert creativity without collapsing tacit practices into overly rationalized structures. This is less about optimizing usability than about sustaining diversity and preventing any single level of expertise from being normatively imposed as the measure of creative legitimacy.

Looking forward, three directions are particularly pressing. First, longitudinal studies could examine whether redistributive effects are temporary or whether they shift the longer-term politics of authorship in human--AI co-creation. Second, integrating methods such as eye-tracking or cognitive load analysis could help trace how users negotiate control and attention when engaging with reasoning surfaces, exposing subtle dynamics of domination or resistance. Third, moving into multi-user scenarios may surface how shared reasoning media enable collective creativity while also amplifying new conflicts over authorship, legitimacy, and authority. Taken together, these directions push future research beyond system refinement toward deeper critical engagement with how design choices configure whose voices are amplified, whose practices are disciplined, and whose creativity is recognized as legitimate.

\section{Conclusion}
This paper reframes the challenge of human–AI co-creation by centering authority in the knowing–doing cycle. We advanced the first systematic framework for redistributing authority in human-AI co-creation through three anti-oppressive design principles: contestability, agency, and plurality. Implemented in a minimally sufficient prototype and evaluated with 180 music practitioners across levels of expertise, these principles enabled users to contest hidden assumptions, redirect generative processes, and defend minority aesthetics. The findings demonstrate that authority over the passage from knowing to doing is not fixed but redistributable, transforming creative collaboration from a model-controlled pipeline into a negotiated medium.

Beyond music, our work establishes a foundation for rethinking creativity-support systems as sites where authority is made visible, contestable, and shareable. By foregrounding justice over efficiency and power over performance, we contribute a new paradigm for critical computing, one in which human–AI co-creation is not only about producing outputs but about redistributing the conditions of creative legitimacy.


\bibliographystyle{ACM-Reference-Format}
\bibliography{sample-base}


\begin{thebibliography}{91}


\ifx \showCODEN    \undefined \def \showCODEN     #1{\unskip}     \fi
\ifx \showISBNx    \undefined \def \showISBNx     #1{\unskip}     \fi
\ifx \showISBNxiii \undefined \def \showISBNxiii  #1{\unskip}     \fi
\ifx \showISSN     \undefined \def \showISSN      #1{\unskip}     \fi
\ifx \showLCCN     \undefined \def \showLCCN      #1{\unskip}     \fi
\ifx \shownote     \undefined \def \shownote      #1{#1}          \fi
\ifx \showarticletitle \undefined \def \showarticletitle #1{#1}   \fi
\ifx \showURL      \undefined \def \showURL       {\relax}        \fi
\providecommand\bibfield[2]{#2}
\providecommand\bibinfo[2]{#2}
\providecommand\natexlab[1]{#1}
\providecommand\showeprint[2][]{arXiv:#2}

\bibitem[Arawjo et~al\mbox{.}(2024)]%
        {arawjo2024chainforge}
\bibfield{author}{\bibinfo{person}{Ian Arawjo}, \bibinfo{person}{Chelse Swoopes}, \bibinfo{person}{Priyan Vaithilingam}, \bibinfo{person}{Martin Wattenberg}, {and} \bibinfo{person}{Elena~L Glassman}.} \bibinfo{year}{2024}\natexlab{}.
\newblock \showarticletitle{Chainforge: A visual toolkit for prompt engineering and llm hypothesis testing}. In \bibinfo{booktitle}{\emph{Proceedings of the 2024 CHI Conference on Human Factors in Computing Systems}}. \bibinfo{pages}{1--18}.
\newblock


\bibitem[Bardzell(2010)]%
        {bardzell2010feminist}
\bibfield{author}{\bibinfo{person}{Shaowen Bardzell}.} \bibinfo{year}{2010}\natexlab{}.
\newblock \showarticletitle{Feminist HCI: taking stock and outlining an agenda for design}. In \bibinfo{booktitle}{\emph{Proceedings of the SIGCHI conference on human factors in computing systems}}. \bibinfo{pages}{1301--1310}.
\newblock


\bibitem[Besta et~al\mbox{.}(2024)]%
        {besta2024graphofthoughts}
\bibfield{author}{\bibinfo{person}{Maciej Besta}, \bibinfo{person}{Torsten Hoefler}, {and} \bibinfo{person}{et al.}} \bibinfo{year}{2024}\natexlab{}.
\newblock \showarticletitle{Graph of Thoughts: Solving Elaborate Problems with Large Language Models}. In \bibinfo{booktitle}{\emph{AAAI}}.
\newblock
\urldef\tempurl%
\url{https://ojs.aaai.org/index.php/AAAI/article/view/29918}
\showURL{%
\tempurl}


\bibitem[Bourdieu(2018)]%
        {bourdieu2018distinction}
\bibfield{author}{\bibinfo{person}{Pierre Bourdieu}.} \bibinfo{year}{2018}\natexlab{}.
\newblock \showarticletitle{Distinction a social critique of the judgement of taste}.
\newblock In \bibinfo{booktitle}{\emph{Inequality}}. \bibinfo{publisher}{Routledge}, \bibinfo{pages}{287--318}.
\newblock


\bibitem[Cavez et~al\mbox{.}(2025)]%
        {cavez2025euterpen}
\bibfield{author}{\bibinfo{person}{Vincent Cavez}, \bibinfo{person}{Catherine Letondal}, \bibinfo{person}{Caroline Appert}, {and} \bibinfo{person}{Emmanuel Pietriga}.} \bibinfo{year}{2025}\natexlab{}.
\newblock \showarticletitle{EuterPen: Unleashing Creative Expression in Music Score Writing}. In \bibinfo{booktitle}{\emph{Proceedings of the 2025 CHI Conference on Human Factors in Computing Systems}}. \bibinfo{pages}{1--16}.
\newblock


\bibitem[Choi et~al\mbox{.}(2025b)]%
        {Expandora}
\bibfield{author}{\bibinfo{person}{DaEun Choi}, \bibinfo{person}{Kihoon Son}, \bibinfo{person}{Hyunjoon Jung}, {and} \bibinfo{person}{Juho Kim}.} \bibinfo{year}{2025}\natexlab{b}.
\newblock \showarticletitle{Expandora: Broadening Design Exploration with Text-to-Image Model}. In \bibinfo{booktitle}{\emph{Extended Abstracts of the 2025 CHI Conference on Human Factors in Computing Systems (CHI EA ’25)}}. \bibinfo{publisher}{ACM}.
\newblock
\href{https://doi.org/10.1145/3706599.3720189}{doi:\nolinkurl{10.1145/3706599.3720189}}


\bibitem[Choi et~al\mbox{.}(2025a)]%
        {choi2025understanding}
\bibfield{author}{\bibinfo{person}{Youjin Choi}, \bibinfo{person}{JaeYoung Moon}, \bibinfo{person}{JinYoung Yoo}, {and} \bibinfo{person}{Jin-Hyuk Hong}.} \bibinfo{year}{2025}\natexlab{a}.
\newblock \showarticletitle{Understanding the Potentials and Limitations of Prompt-based Music Generative AI}. In \bibinfo{booktitle}{\emph{Proceedings of the 2025 CHI Conference on Human Factors in Computing Systems}}. \bibinfo{pages}{1--15}.
\newblock


\bibitem[Choi et~al\mbox{.}(2023)]%
        {choi2023toward}
\bibfield{author}{\bibinfo{person}{Yunjae~J Choi}, \bibinfo{person}{Minha Lee}, {and} \bibinfo{person}{Sangsu Lee}.} \bibinfo{year}{2023}\natexlab{}.
\newblock \showarticletitle{Toward a multilingual conversational agent: Challenges and expectations of code-mixing multilingual users}. In \bibinfo{booktitle}{\emph{Proceedings of the 2023 CHI conference on human factors in computing systems}}. \bibinfo{pages}{1--17}.
\newblock


\bibitem[Costanza-Chock(2020)]%
        {costanza2020design}
\bibfield{author}{\bibinfo{person}{Sasha Costanza-Chock}.} \bibinfo{year}{2020}\natexlab{}.
\newblock \bibinfo{booktitle}{\emph{Design justice: Community-led practices to build the worlds we need}}.
\newblock \bibinfo{publisher}{The MIT Press}.
\newblock


\bibitem[Crisp(2014)]%
        {crisp2014aristotle}
\bibfield{author}{\bibinfo{person}{Roger Crisp}.} \bibinfo{year}{2014}\natexlab{}.
\newblock \bibinfo{booktitle}{\emph{Aristotle: nicomachean ethics}}.
\newblock \bibinfo{publisher}{Cambridge University Press}.
\newblock


\bibitem[Dang et~al\mbox{.}(2022)]%
        {dang2022ganslider}
\bibfield{author}{\bibinfo{person}{Hai Dang}, \bibinfo{person}{Lukas Mecke}, {and} \bibinfo{person}{Daniel Buschek}.} \bibinfo{year}{2022}\natexlab{}.
\newblock \showarticletitle{Ganslider: How users control generative models for images using multiple sliders with and without feedforward information}. In \bibinfo{booktitle}{\emph{Proceedings of the 2022 CHI Conference on Human Factors in Computing Systems}}. \bibinfo{pages}{1--15}.
\newblock


\bibitem[Davis and Rafner(2025)]%
        {davis2025ai}
\bibfield{author}{\bibinfo{person}{Nicholas Davis} {and} \bibinfo{person}{Janet Rafner}.} \bibinfo{year}{2025}\natexlab{}.
\newblock \showarticletitle{AI Drawing Partner: Co-Creative Drawing Agent and Research Platform to Model Co-Creation}.
\newblock \bibinfo{journal}{\emph{arXiv preprint arXiv:2501.06607}} (\bibinfo{year}{2025}).
\newblock


\bibitem[Dewey(2024)]%
        {dewey2024art}
\bibfield{author}{\bibinfo{person}{John Dewey}.} \bibinfo{year}{2024}\natexlab{}.
\newblock \showarticletitle{Art as experience}.
\newblock In \bibinfo{booktitle}{\emph{Anthropology of the Arts}}. \bibinfo{publisher}{Routledge}, \bibinfo{pages}{37--45}.
\newblock


\bibitem[Dhillon et~al\mbox{.}(2024)]%
        {Dhillon2024}
\bibfield{author}{\bibinfo{person}{Paramveer~S. Dhillon}, \bibinfo{person}{Somayeh Molaei}, \bibinfo{person}{Jiaqi Li}, \bibinfo{person}{Maximilian Golub}, \bibinfo{person}{Shaochun Zheng}, {and} \bibinfo{person}{Lionel~P. Robert}.} \bibinfo{year}{2024}\natexlab{}.
\newblock \showarticletitle{Shaping Human–AI Collaboration: Varied Scaffolding Levels in Co-Writing with Language Models}. In \bibinfo{booktitle}{\emph{Proceedings of the 2024 CHI Conference on Human Factors in Computing Systems}} \emph{(\bibinfo{series}{CHI '24})}. \bibinfo{publisher}{ACM}.
\newblock
\href{https://doi.org/10.1145/3613904.3642134}{doi:\nolinkurl{10.1145/3613904.3642134}}


\bibitem[Ehsan et~al\mbox{.}(2021)]%
        {ehsan2021expanding}
\bibfield{author}{\bibinfo{person}{Upol Ehsan}, \bibinfo{person}{Q~Vera Liao}, \bibinfo{person}{Michael Muller}, \bibinfo{person}{Mark~O Riedl}, {and} \bibinfo{person}{Justin~D Weisz}.} \bibinfo{year}{2021}\natexlab{}.
\newblock \showarticletitle{Expanding explainability: Towards social transparency in ai systems}. In \bibinfo{booktitle}{\emph{Proceedings of the 2021 CHI conference on human factors in computing systems}}. \bibinfo{pages}{1--19}.
\newblock


\bibitem[Epperson et~al\mbox{.}(2025)]%
        {epperson2025interactive}
\bibfield{author}{\bibinfo{person}{Will Epperson}, \bibinfo{person}{Gagan Bansal}, \bibinfo{person}{Victor~C Dibia}, \bibinfo{person}{Adam Fourney}, \bibinfo{person}{Jack Gerrits}, \bibinfo{person}{Erkang Zhu}, {and} \bibinfo{person}{Saleema Amershi}.} \bibinfo{year}{2025}\natexlab{}.
\newblock \showarticletitle{Interactive debugging and steering of multi-agent ai systems}. In \bibinfo{booktitle}{\emph{Proceedings of the 2025 CHI Conference on Human Factors in Computing Systems}}. \bibinfo{pages}{1--15}.
\newblock


\bibitem[Eslami et~al\mbox{.}(2016)]%
        {eslami2016first}
\bibfield{author}{\bibinfo{person}{Motahhare Eslami}, \bibinfo{person}{Karrie Karahalios}, \bibinfo{person}{Christian Sandvig}, \bibinfo{person}{Kristen Vaccaro}, \bibinfo{person}{Aimee Rickman}, \bibinfo{person}{Kevin Hamilton}, {and} \bibinfo{person}{Alex Kirlik}.} \bibinfo{year}{2016}\natexlab{}.
\newblock \showarticletitle{First I" like" it, then I hide it: Folk Theories of Social Feeds}. In \bibinfo{booktitle}{\emph{Proceedings of the 2016 cHI conference on human factors in computing systems}}. \bibinfo{pages}{2371--2382}.
\newblock


\bibitem[Everett(1999)]%
        {everett1999beatles}
\bibfield{author}{\bibinfo{person}{Walter Everett}.} \bibinfo{year}{1999}\natexlab{}.
\newblock \bibinfo{booktitle}{\emph{The Beatles as musicians: Revolver through the Anthology}}.
\newblock \bibinfo{publisher}{Oxford University Press, USA}.
\newblock


\bibitem[Evirgen and Chen(2023)]%
        {evirgen2023ganravel}
\bibfield{author}{\bibinfo{person}{Noyan Evirgen} {and} \bibinfo{person}{Xiang'Anthony Chen}.} \bibinfo{year}{2023}\natexlab{}.
\newblock \showarticletitle{Ganravel: User-driven direction disentanglement in generative adversarial networks}. In \bibinfo{booktitle}{\emph{Proceedings of the 2023 CHI Conference on Human Factors in Computing Systems}}. \bibinfo{pages}{1--15}.
\newblock


\bibitem[Foucault(2020)]%
        {foucault2020power}
\bibfield{author}{\bibinfo{person}{Michel Foucault}.} \bibinfo{year}{2020}\natexlab{}.
\newblock \showarticletitle{Power/knowledge}.
\newblock In \bibinfo{booktitle}{\emph{The new social theory reader}}. \bibinfo{publisher}{Routledge}, \bibinfo{pages}{73--79}.
\newblock


\bibitem[Gero and Liu(2024)]%
        {gero2024sensemaking}
\bibfield{author}{\bibinfo{person}{Kig Gero} {and} \bibinfo{person}{et~al. Liu}.} \bibinfo{year}{2024}\natexlab{}.
\newblock \showarticletitle{Supporting Sensemaking of Large Language Model Outputs at Scale}. In \bibinfo{booktitle}{\emph{CHI}}.
\newblock
\urldef\tempurl%
\url{https://dl.acm.org/doi/10.1145/3613904.3642139}
\showURL{%
\tempurl}


\bibitem[{GPT-4o}(2025)]%
        {gpt4o}
\bibfield{author}{\bibinfo{person}{{GPT-4o}}.} \bibinfo{year}{2025}\natexlab{}.
\newblock \bibinfo{title}{Official website}.
\newblock \bibinfo{howpublished}{\url{https://chatgpt.com/}}.
\newblock
\newblock
\shownote{Accessed: 2025-09-01}.


\bibitem[Green and Chen(2019)]%
        {green2019disparate}
\bibfield{author}{\bibinfo{person}{Ben Green} {and} \bibinfo{person}{Yiling Chen}.} \bibinfo{year}{2019}\natexlab{}.
\newblock \showarticletitle{Disparate interactions: An algorithm-in-the-loop analysis of fairness in risk assessments}. In \bibinfo{booktitle}{\emph{Proceedings of the conference on fairness, accountability, and transparency}}. \bibinfo{pages}{90--99}.
\newblock


\bibitem[He et~al\mbox{.}(2025)]%
        {He2025}
\bibfield{author}{\bibinfo{person}{Jessica He}, \bibinfo{person}{Stephanie Houde}, {and} \bibinfo{person}{Justin~D. Weisz}.} \bibinfo{year}{2025}\natexlab{}.
\newblock \showarticletitle{Which Contributions Deserve Credit? Perceptions of Attribution in Human–AI Co-Creation}. In \bibinfo{booktitle}{\emph{Proceedings of the 2025 CHI Conference on Human Factors in Computing Systems}} \emph{(\bibinfo{series}{CHI '25})}. \bibinfo{publisher}{ACM}.
\newblock
\href{https://doi.org/10.1145/3706598.3713522}{doi:\nolinkurl{10.1145/3706598.3713522}}


\bibitem[Hyde(1985)]%
        {hyde1985musical}
\bibfield{author}{\bibinfo{person}{Martha~M Hyde}.} \bibinfo{year}{1985}\natexlab{}.
\newblock \showarticletitle{Musical Form and the Development of Schoenberg's" Twelve-Tone Method"}.
\newblock \bibinfo{journal}{\emph{Journal of Music Theory}} \bibinfo{volume}{29}, \bibinfo{number}{1} (\bibinfo{year}{1985}), \bibinfo{pages}{85--143}.
\newblock


\bibitem[Jacovi et~al\mbox{.}(2024)]%
        {jacovi2024reveal}
\bibfield{author}{\bibinfo{person}{Alon Jacovi}, \bibinfo{person}{Gregor Geigle}, {and} \bibinfo{person}{et al.}} \bibinfo{year}{2024}\natexlab{}.
\newblock \showarticletitle{{REVEAL}: A Benchmark for Verifiers of Reasoning Chains}. In \bibinfo{booktitle}{\emph{ACL}}.
\newblock
\urldef\tempurl%
\url{https://aclanthology.org/2024.acl-long.254/}
\showURL{%
\tempurl}


\bibitem[Jain et~al\mbox{.}(2025)]%
        {AdaptiveSliders}
\bibfield{author}{\bibinfo{person}{Rahul Jain}, \bibinfo{person}{Amit Goel}, \bibinfo{person}{Koichiro Niinuma}, {and} \bibinfo{person}{Aakar Gupta}.} \bibinfo{year}{2025}\natexlab{}.
\newblock \showarticletitle{AdaptiveSliders: User-Aligned Semantic Slider-based Editing of Text-to-Image Model Output}. In \bibinfo{booktitle}{\emph{Proceedings of the 2025 CHI Conference on Human Factors in Computing Systems (CHI ’25)}}. \bibinfo{publisher}{ACM}.
\newblock
\href{https://doi.org/10.1145/3706598.3713860}{doi:\nolinkurl{10.1145/3706598.3713860}}


\bibitem[Jakesch et~al\mbox{.}(2023)]%
        {Jakesch2023}
\bibfield{author}{\bibinfo{person}{Maurice Jakesch}, \bibinfo{person}{Advait Bhat}, \bibinfo{person}{Daniel Buschek}, \bibinfo{person}{Lior Zalmanson}, {and} \bibinfo{person}{Mor Naaman}.} \bibinfo{year}{2023}\natexlab{}.
\newblock \showarticletitle{Co-Writing with Opinionated Language Models Affects Users' Views}. In \bibinfo{booktitle}{\emph{Proceedings of the 2023 CHI Conference on Human Factors in Computing Systems}} \emph{(\bibinfo{series}{CHI '23})}. \bibinfo{publisher}{ACM}.
\newblock
\href{https://doi.org/10.1145/3544548.3581196}{doi:\nolinkurl{10.1145/3544548.3581196}}


\bibitem[Kahng and et~al.(2024)]%
        {kahng2024llmcomparator}
\bibfield{author}{\bibinfo{person}{Minsuk Kahng} {and} \bibinfo{person}{et al.}} \bibinfo{year}{2024}\natexlab{}.
\newblock \showarticletitle{{LLM Comparator}: Visual Analytics for Side-by-Side Evaluation of Large Language Models}. In \bibinfo{booktitle}{\emph{CHI}}.
\newblock
\urldef\tempurl%
\url{https://arxiv.org/abs/2402.10524}
\showURL{%
\tempurl}


\bibitem[Kant(2000)]%
        {kant2000critique}
\bibfield{author}{\bibinfo{person}{Immanuel Kant}.} \bibinfo{year}{2000}\natexlab{}.
\newblock \bibinfo{booktitle}{\emph{Critique of the Power of Judgment}}.
\newblock \bibinfo{publisher}{Cambridge University Press}.
\newblock


\bibitem[Kapania et~al\mbox{.}(2023)]%
        {kapania2023hunt}
\bibfield{author}{\bibinfo{person}{Shivani Kapania}, \bibinfo{person}{Alex~S Taylor}, {and} \bibinfo{person}{Ding Wang}.} \bibinfo{year}{2023}\natexlab{}.
\newblock \showarticletitle{A hunt for the snark: Annotator diversity in data practices}. In \bibinfo{booktitle}{\emph{Proceedings of the 2023 CHI Conference on Human Factors in Computing Systems}}. \bibinfo{pages}{1--15}.
\newblock


\bibitem[Khan et~al\mbox{.}(2025)]%
        {Khan2025}
\bibfield{author}{\bibinfo{person}{Abidullah Khan}, \bibinfo{person}{Atefeh Shokrizadeh}, {and} \bibinfo{person}{Jinghui Cheng}.} \bibinfo{year}{2025}\natexlab{}.
\newblock \showarticletitle{Beyond Automation: How UI/UX Designers Perceive AI as a Creative Partner in the Divergent Thinking Stages}. In \bibinfo{booktitle}{\emph{Proceedings of the 2025 CHI Conference on Human Factors in Computing Systems}} \emph{(\bibinfo{series}{CHI '25})}. \bibinfo{publisher}{ACM}.
\newblock


\bibitem[Kim et~al\mbox{.}(2025)]%
        {Kim2025}
\bibfield{author}{\bibinfo{person}{Chae-Young Kim}, \bibinfo{person}{Soojeong Kim}, \bibinfo{person}{Jong-Yun Jang}, {and} \bibinfo{person}{Sang-Hyeok Choi}.} \bibinfo{year}{2025}\natexlab{}.
\newblock \showarticletitle{Bridging Generations Using AI-Supported Co-Creative Activities}. In \bibinfo{booktitle}{\emph{Proceedings of the 2025 CHI Conference on Human Factors in Computing Systems}} \emph{(\bibinfo{series}{CHI '25})}. \bibinfo{publisher}{ACM}.
\newblock
\href{https://doi.org/10.1145/3706598.3713718}{doi:\nolinkurl{10.1145/3706598.3713718}}


\bibitem[Kim et~al\mbox{.}(2024)]%
        {KimTaeSoo2024}
\bibfield{author}{\bibinfo{person}{Tae~Soo Kim}, \bibinfo{person}{Yoonjoo Lee}, \bibinfo{person}{Jamin Shin}, \bibinfo{person}{Young-Ho Kim}, {and} \bibinfo{person}{Juho Kim}.} \bibinfo{year}{2024}\natexlab{}.
\newblock \showarticletitle{EvalLM: Interactive Evaluation of Large Language Model Prompts on User-Defined Criteria}. In \bibinfo{booktitle}{\emph{Proceedings of the 2024 {CHI} Conference on Human Factors in Computing Systems}} \emph{(\bibinfo{series}{CHI '24})}. \bibinfo{publisher}{ACM}.
\newblock
\href{https://doi.org/10.1145/3613904.3642216}{doi:\nolinkurl{10.1145/3613904.3642216}}


\bibitem[Kollig et~al\mbox{.}(2025)]%
        {kollig2025fictional}
\bibfield{author}{\bibinfo{person}{Faye Kollig}, \bibinfo{person}{Jessica Pater}, \bibinfo{person}{Fayika~Farhat Nova}, {and} \bibinfo{person}{Casey Fiesler}.} \bibinfo{year}{2025}\natexlab{}.
\newblock \showarticletitle{Fictional Failures and Real-World Lessons: Ethical Speculation Through Design Fiction on Emotional Support Conversational AI}. In \bibinfo{booktitle}{\emph{Proceedings of the 2025 CHI Conference on Human Factors in Computing Systems}}. \bibinfo{pages}{1--15}.
\newblock


\bibitem[Kopeinik et~al\mbox{.}(2023)]%
        {kopeinik2023show}
\bibfield{author}{\bibinfo{person}{Simone Kopeinik}, \bibinfo{person}{Martina Mara}, \bibinfo{person}{Linda Ratz}, \bibinfo{person}{Klara Krieg}, \bibinfo{person}{Markus Schedl}, {and} \bibinfo{person}{Navid Rekabsaz}.} \bibinfo{year}{2023}\natexlab{}.
\newblock \showarticletitle{Show me a" male nurse"! how gender bias is reflected in the query formulation of search engine users}. In \bibinfo{booktitle}{\emph{Proceedings of the 2023 CHI Conference on Human Factors in Computing Systems}}. \bibinfo{pages}{1--15}.
\newblock


\bibitem[Krol et~al\mbox{.}(2025)]%
        {krol2025exploring}
\bibfield{author}{\bibinfo{person}{Stephen~James Krol}, \bibinfo{person}{Maria~Teresa Llano~Rodriguez}, {and} \bibinfo{person}{Miguel~J Loor~Paredes}.} \bibinfo{year}{2025}\natexlab{}.
\newblock \showarticletitle{Exploring the Needs of Practising Musicians in Co-Creative AI Through Co-Design}. In \bibinfo{booktitle}{\emph{Proceedings of the 2025 CHI Conference on Human Factors in Computing Systems}}. \bibinfo{pages}{1--13}.
\newblock


\bibitem[Lee et~al\mbox{.}(2025)]%
        {lee2025mvprompt}
\bibfield{author}{\bibinfo{person}{ChungHa Lee}, \bibinfo{person}{DaeHo Lee}, {and} \bibinfo{person}{Jin-Hyuk Hong}.} \bibinfo{year}{2025}\natexlab{}.
\newblock \showarticletitle{MVPrompt: Building Music-Visual Prompts for AI Artists to Craft Music Video Mise-en-sc{\`e}ne}. In \bibinfo{booktitle}{\emph{Proceedings of the 2025 CHI Conference on Human Factors in Computing Systems}}. \bibinfo{pages}{1--21}.
\newblock


\bibitem[Lee et~al\mbox{.}(2019)]%
        {lee2019webuildai}
\bibfield{author}{\bibinfo{person}{Min~Kyung Lee}, \bibinfo{person}{Daniel Kusbit}, \bibinfo{person}{Anson Kahng}, \bibinfo{person}{Ji~Tae Kim}, \bibinfo{person}{Xinran Yuan}, \bibinfo{person}{Allissa Chan}, \bibinfo{person}{Daniel See}, \bibinfo{person}{Ritesh Noothigattu}, \bibinfo{person}{Siheon Lee}, \bibinfo{person}{Alexandros Psomas}, {et~al\mbox{.}}} \bibinfo{year}{2019}\natexlab{}.
\newblock \showarticletitle{WeBuildAI: Participatory framework for algorithmic governance}.
\newblock \bibinfo{journal}{\emph{Proceedings of the ACM on human-computer interaction}} \bibinfo{volume}{3}, \bibinfo{number}{CSCW} (\bibinfo{year}{2019}), \bibinfo{pages}{1--35}.
\newblock


\bibitem[Lewicki et~al\mbox{.}(2023)]%
        {lewicki2023out}
\bibfield{author}{\bibinfo{person}{Kornel Lewicki}, \bibinfo{person}{Michelle Seng~Ah Lee}, \bibinfo{person}{Jennifer Cobbe}, {and} \bibinfo{person}{Jatinder Singh}.} \bibinfo{year}{2023}\natexlab{}.
\newblock \showarticletitle{Out of context: Investigating the bias and fairness concerns of “artificial intelligence as a service”}. In \bibinfo{booktitle}{\emph{Proceedings of the 2023 CHI Conference on Human Factors in Computing Systems}}. \bibinfo{pages}{1--17}.
\newblock


\bibitem[Lin et~al\mbox{.}(2025)]%
        {SketchFlex}
\bibfield{author}{\bibinfo{person}{Haichuan Lin}, \bibinfo{person}{Yilin Ye}, \bibinfo{person}{Jiazhi Xia}, {and} \bibinfo{person}{Wei Zeng}.} \bibinfo{year}{2025}\natexlab{}.
\newblock \showarticletitle{SketchFlex: Facilitating Spatial--Semantic Coherence in Text-to-Image Generation with Region-Based Sketches}. In \bibinfo{booktitle}{\emph{Proceedings of the 2025 CHI Conference on Human Factors in Computing Systems (CHI ’25)}}. \bibinfo{publisher}{ACM}.
\newblock
\href{https://doi.org/10.1145/3706598.3713801}{doi:\nolinkurl{10.1145/3706598.3713801}}


\bibitem[Lin et~al\mbox{.}(2024)]%
        {Lin2024}
\bibfield{author}{\bibinfo{person}{Susan Lin}, \bibinfo{person}{Jeremy Warner}, \bibinfo{person}{J.~D. Zamfirescu-Pereira}, \bibinfo{person}{Matthew~G. Lee}, \bibinfo{person}{Sauhard Jain}, \bibinfo{person}{Shanqing Cai}, \bibinfo{person}{Piyawat Lertvittayakumjorn}, \bibinfo{person}{Michael Huang}, \bibinfo{person}{Shumin Zhai}, \bibinfo{person}{Bj{\"o}rn Hartmann}, {and} \bibinfo{person}{Can Liu}.} \bibinfo{year}{2024}\natexlab{}.
\newblock \showarticletitle{Rambler: Supporting Writing with Speech via {LLM}-Assisted Gist Manipulation}. In \bibinfo{booktitle}{\emph{Proceedings of the 2024 CHI Conference on Human Factors in Computing Systems}} \emph{(\bibinfo{series}{CHI '24})}. \bibinfo{publisher}{ACM}.
\newblock
\href{https://doi.org/10.1145/3613904.3642217}{doi:\nolinkurl{10.1145/3613904.3642217}}


\bibitem[Liu(2023)]%
        {Liu2023}
\bibfield{author}{\bibinfo{person}{Vivian Liu}.} \bibinfo{year}{2023}\natexlab{}.
\newblock \showarticletitle{Beyond Text-to-Image: Multimodal Prompts to Explore Generative AI}. In \bibinfo{booktitle}{\emph{Extended Abstracts of the 2023 CHI Conference on Human Factors in Computing Systems}} \emph{(\bibinfo{series}{CHI EA '23})}. \bibinfo{publisher}{ACM}, \bibinfo{pages}{482:1--482:6}.
\newblock
\href{https://doi.org/10.1145/3544549.3577043}{doi:\nolinkurl{10.1145/3544549.3577043}}


\bibitem[Liu and Chilton(2022)]%
        {liu2022design}
\bibfield{author}{\bibinfo{person}{Vivian Liu} {and} \bibinfo{person}{Lydia~B Chilton}.} \bibinfo{year}{2022}\natexlab{}.
\newblock \showarticletitle{Design guidelines for prompt engineering text-to-image generative models}. In \bibinfo{booktitle}{\emph{Proceedings of the 2022 CHI conference on human factors in computing systems}}. \bibinfo{pages}{1--23}.
\newblock


\bibitem[Lyons et~al\mbox{.}(2021)]%
        {lyons2021conceptualising}
\bibfield{author}{\bibinfo{person}{Henrietta Lyons}, \bibinfo{person}{Eduardo Velloso}, {and} \bibinfo{person}{Tim Miller}.} \bibinfo{year}{2021}\natexlab{}.
\newblock \showarticletitle{Conceptualising contestability: Perspectives on contesting algorithmic decisions}.
\newblock \bibinfo{journal}{\emph{Proceedings of the ACM on Human-Computer Interaction}} \bibinfo{volume}{5}, \bibinfo{number}{CSCW1} (\bibinfo{year}{2021}), \bibinfo{pages}{1--25}.
\newblock


\bibitem[Mahdavi~Goloujeh et~al\mbox{.}(2024a)]%
        {mahdavi2024ai}
\bibfield{author}{\bibinfo{person}{Atefeh Mahdavi~Goloujeh}, \bibinfo{person}{Anne Sullivan}, {and} \bibinfo{person}{Brian Magerko}.} \bibinfo{year}{2024}\natexlab{a}.
\newblock \showarticletitle{Is it AI or is it me? Understanding users’ prompt journey with text-to-image generative AI tools}. In \bibinfo{booktitle}{\emph{Proceedings of the 2024 CHI Conference on Human Factors in Computing Systems}}. \bibinfo{pages}{1--13}.
\newblock


\bibitem[Mahdavi~Goloujeh et~al\mbox{.}(2024b)]%
        {GoloujehEA2024}
\bibfield{author}{\bibinfo{person}{Atefeh Mahdavi~Goloujeh}, \bibinfo{person}{Anne~M. Sullivan}, {and} \bibinfo{person}{Brian Magerko}.} \bibinfo{year}{2024}\natexlab{b}.
\newblock \showarticletitle{The Social Construction of Generative AI Prompts}. In \bibinfo{booktitle}{\emph{Extended Abstracts of the 2024 CHI Conference on Human Factors in Computing Systems}} \emph{(\bibinfo{series}{CHI EA '24})}. \bibinfo{publisher}{ACM}, \bibinfo{pages}{320:1--320:7}.
\newblock
\href{https://doi.org/10.1145/3613905.3650947}{doi:\nolinkurl{10.1145/3613905.3650947}}


\bibitem[Malandro(2024)]%
        {malandro2024composer}
\bibfield{author}{\bibinfo{person}{Martin~E Malandro}.} \bibinfo{year}{2024}\natexlab{}.
\newblock \showarticletitle{Composer's Assistant 2: Interactive Multi-Track MIDI Infilling with Fine-Grained User Control}.
\newblock \bibinfo{journal}{\emph{arXiv preprint arXiv:2407.14700}} (\bibinfo{year}{2024}).
\newblock


\bibitem[Marcus(2016)]%
        {marcus2016schoenberg}
\bibfield{author}{\bibinfo{person}{Kenneth~H Marcus}.} \bibinfo{year}{2016}\natexlab{}.
\newblock \bibinfo{booktitle}{\emph{Schoenberg and Hollywood Modernism}}.
\newblock \bibinfo{publisher}{Cambridge University Press}.
\newblock


\bibitem[Masson et~al\mbox{.}(2024)]%
        {DirectGPT}
\bibfield{author}{\bibinfo{person}{Damien Masson}, \bibinfo{person}{Sylvain Malacria}, \bibinfo{person}{G\'ery Casiez}, {and} \bibinfo{person}{Daniel Vogel}.} \bibinfo{year}{2024}\natexlab{}.
\newblock \showarticletitle{DirectGPT: A Direct Manipulation Interface to Interact with Large Language Models}. In \bibinfo{booktitle}{\emph{Proceedings of the 2024 CHI Conference on Human Factors in Computing Systems (CHI ’24)}}. \bibinfo{publisher}{ACM}.
\newblock
\href{https://doi.org/10.1145/3613904.3642462}{doi:\nolinkurl{10.1145/3613904.3642462}}


\bibitem[Meyer(1954)]%
        {meyer1954emotion}
\bibfield{author}{\bibinfo{person}{Leonard~B Meyer}.} \bibinfo{year}{1954}\natexlab{}.
\newblock \emph{\bibinfo{title}{Emotion and meaning in music}}.
\newblock \bibinfo{thesistype}{Ph.\,D. Dissertation}. \bibinfo{school}{The University of Chicago}.
\newblock


\bibitem[{Midjourney}(2025)]%
        {midjourney2024}
\bibfield{author}{\bibinfo{person}{{Midjourney}}.} \bibinfo{year}{2025}\natexlab{}.
\newblock \bibinfo{title}{Official website}.
\newblock \bibinfo{howpublished}{\url{https://www.midjourney.com/}}.
\newblock
\newblock
\shownote{Accessed: 2025-08-01}.


\bibitem[Pan et~al\mbox{.}(2023)]%
        {pan2023drag}
\bibfield{author}{\bibinfo{person}{Xingang Pan}, \bibinfo{person}{Ayush Tewari}, \bibinfo{person}{Thomas Leimk{\"u}hler}, \bibinfo{person}{Lingjie Liu}, \bibinfo{person}{Abhimitra Meka}, {and} \bibinfo{person}{Christian Theobalt}.} \bibinfo{year}{2023}\natexlab{}.
\newblock \showarticletitle{Drag your gan: Interactive point-based manipulation on the generative image manifold}. In \bibinfo{booktitle}{\emph{ACM SIGGRAPH 2023 conference proceedings}}. \bibinfo{pages}{1--11}.
\newblock


\bibitem[Pang et~al\mbox{.}(2025)]%
        {Pang2025}
\bibfield{author}{\bibinfo{person}{Rock~Yuren Pang}, \bibinfo{person}{K.~J.~Kevin Feng}, \bibinfo{person}{Shangbin Feng}, \bibinfo{person}{Chu Li}, \bibinfo{person}{Weijia Shi}, \bibinfo{person}{Yulia Tsvetkov}, \bibinfo{person}{Jeffrey Heer}, {and} \bibinfo{person}{Katharina Reinecke}.} \bibinfo{year}{2025}\natexlab{}.
\newblock \showarticletitle{Interactive Reasoning: Visualizing and Controlling Chain-of-Thought Reasoning in Large Language Models}. In \bibinfo{booktitle}{\emph{Proceedings of the 2025 {CHI} Conference on Human Factors in Computing Systems}} \emph{(\bibinfo{series}{CHI '25})}. \bibinfo{publisher}{ACM}.
\newblock
\href{https://doi.org/10.1145/3706598.3714020}{doi:\nolinkurl{10.1145/3706598.3714020}}


\bibitem[Peng et~al\mbox{.}(2025)]%
        {FusAIn}
\bibfield{author}{\bibinfo{person}{Xiaohan Peng}, \bibinfo{person}{Janin Koch}, {and} \bibinfo{person}{Wendy~E. Mackay}.} \bibinfo{year}{2025}\natexlab{}.
\newblock \showarticletitle{FusAIn: Composing Generative AI Visual Prompts Using Pen-based Interaction}. In \bibinfo{booktitle}{\emph{Proceedings of the 2025 CHI Conference on Human Factors in Computing Systems (CHI ’25)}}. \bibinfo{publisher}{ACM}.
\newblock
\href{https://doi.org/10.1145/3706598.3714027}{doi:\nolinkurl{10.1145/3706598.3714027}}


\bibitem[Porquet et~al\mbox{.}(2025)]%
        {porquet2025copying}
\bibfield{author}{\bibinfo{person}{Julien Porquet}, \bibinfo{person}{Sitong Wang}, {and} \bibinfo{person}{Lydia~B Chilton}.} \bibinfo{year}{2025}\natexlab{}.
\newblock \showarticletitle{Copying style, Extracting value: Illustrators' Perception of AI Style Transfer and its Impact on Creative Labor}. In \bibinfo{booktitle}{\emph{Proceedings of the 2025 CHI Conference on Human Factors in Computing Systems}}. \bibinfo{pages}{1--16}.
\newblock


\bibitem[Qadri et~al\mbox{.}(2025)]%
        {qadri2025ai}
\bibfield{author}{\bibinfo{person}{Rida Qadri}, \bibinfo{person}{Piotr Mirowski}, {and} \bibinfo{person}{Remi Denton}.} \bibinfo{year}{2025}\natexlab{}.
\newblock \showarticletitle{AI and Non-Western Art Worlds: Reimagining Critical AI Futures through Artistic Inquiry and Situated Dialogue}. In \bibinfo{booktitle}{\emph{Proceedings of the 2025 CHI Conference on Human Factors in Computing Systems}}. \bibinfo{pages}{1--17}.
\newblock


\bibitem[Rajcic et~al\mbox{.}(2024)]%
        {Rajcic2024}
\bibfield{author}{\bibinfo{person}{Nina Rajcic}, \bibinfo{person}{Maria~Teresa Llano~Rodriguez}, {and} \bibinfo{person}{Jon McCormack}.} \bibinfo{year}{2024}\natexlab{}.
\newblock \showarticletitle{Towards a Diffractive Analysis of Prompt-Based Generative AI}. In \bibinfo{booktitle}{\emph{Proceedings of the 2024 CHI Conference on Human Factors in Computing Systems}} \emph{(\bibinfo{series}{CHI '24})}. \bibinfo{publisher}{ACM}, \bibinfo{pages}{844:1--844:15}.
\newblock
\href{https://doi.org/10.1145/3613904.3641971}{doi:\nolinkurl{10.1145/3613904.3641971}}


\bibitem[Ranci{\`e}re(2013)]%
        {ranciere2013politics}
\bibfield{author}{\bibinfo{person}{Jacques Ranci{\`e}re}.} \bibinfo{year}{2013}\natexlab{}.
\newblock \bibinfo{booktitle}{\emph{The politics of aesthetics}}.
\newblock \bibinfo{publisher}{A\&C Black}.
\newblock


\bibitem[Reza et~al\mbox{.}(2024)]%
        {Reza2024}
\bibfield{author}{\bibinfo{person}{Mohi Reza}, \bibinfo{person}{Nathan Laundry}, \bibinfo{person}{Ilya Musabirov}, \bibinfo{person}{Peter Dushniku}, \bibinfo{person}{Zhi Yuan}, \bibinfo{person}{Kashish Mittal}, \bibinfo{person}{Tovi Grossman}, \bibinfo{person}{Michael Liut}, \bibinfo{person}{Anastasia Kuzminykh}, {and} \bibinfo{person}{Joseph~Jay Williams}.} \bibinfo{year}{2024}\natexlab{}.
\newblock \showarticletitle{{ABScribe}: Rapid Exploration \& Organization of Multiple Writing Variations in Human–AI Co-Writing Tasks Using Large Language Models}. In \bibinfo{booktitle}{\emph{Proceedings of the 2024 CHI Conference on Human Factors in Computing Systems}} \emph{(\bibinfo{series}{CHI '24})}. \bibinfo{publisher}{ACM}.
\newblock
\href{https://doi.org/10.1145/3613904.3641899}{doi:\nolinkurl{10.1145/3613904.3641899}}


\bibitem[Schoenberg and Stein(1984)]%
        {schoenberg1984style}
\bibfield{author}{\bibinfo{person}{Arnold Schoenberg} {and} \bibinfo{person}{Leonard Stein}.} \bibinfo{year}{1984}\natexlab{}.
\newblock \bibinfo{booktitle}{\emph{Style and Idea: selected writings of Arnold Schoenberg}}.
\newblock \bibinfo{publisher}{Univ of California Press}.
\newblock


\bibitem[Seaborn et~al\mbox{.}(2023)]%
        {seaborn2023transcending}
\bibfield{author}{\bibinfo{person}{Katie Seaborn}, \bibinfo{person}{Shruti Chandra}, {and} \bibinfo{person}{Thibault Fabre}.} \bibinfo{year}{2023}\natexlab{}.
\newblock \showarticletitle{Transcending the “male code”: implicit masculine biases in NLP contexts}. In \bibinfo{booktitle}{\emph{Proceedings of the 2023 CHI Conference on Human Factors in Computing Systems}}. \bibinfo{pages}{1--19}.
\newblock


\bibitem[Sengers et~al\mbox{.}(2005)]%
        {sengers2005reflective}
\bibfield{author}{\bibinfo{person}{Phoebe Sengers}, \bibinfo{person}{Kirsten Boehner}, \bibinfo{person}{Shay David}, {and} \bibinfo{person}{Joseph'Jofish' Kaye}.} \bibinfo{year}{2005}\natexlab{}.
\newblock \showarticletitle{Reflective design}. In \bibinfo{booktitle}{\emph{Proceedings of the 4th decennial conference on Critical computing: between sense and sensibility}}. \bibinfo{pages}{49--58}.
\newblock


\bibitem[Shao et~al\mbox{.}(2025)]%
        {shao2025unlocking}
\bibfield{author}{\bibinfo{person}{Zekai Shao}, \bibinfo{person}{Siyu Yuan}, \bibinfo{person}{Lin Gao}, \bibinfo{person}{Yixuan He}, \bibinfo{person}{Deqing Yang}, {and} \bibinfo{person}{Siming Chen}.} \bibinfo{year}{2025}\natexlab{}.
\newblock \showarticletitle{Unlocking Scientific Concepts: How Effective Are LLM-Generated Analogies for Student Understanding and Classroom Practice?}. In \bibinfo{booktitle}{\emph{Proceedings of the 2025 CHI Conference on Human Factors in Computing Systems}}. \bibinfo{pages}{1--19}.
\newblock


\bibitem[Shi et~al\mbox{.}(2024)]%
        {shi2024dragdiffusion}
\bibfield{author}{\bibinfo{person}{Yujun Shi}, \bibinfo{person}{Chuhui Xue}, \bibinfo{person}{Jun~Hao Liew}, \bibinfo{person}{Jiachun Pan}, \bibinfo{person}{Hanshu Yan}, \bibinfo{person}{Wenqing Zhang}, \bibinfo{person}{Vincent~YF Tan}, {and} \bibinfo{person}{Song Bai}.} \bibinfo{year}{2024}\natexlab{}.
\newblock \showarticletitle{Dragdiffusion: Harnessing diffusion models for interactive point-based image editing}. In \bibinfo{booktitle}{\emph{Proceedings of the IEEE/CVF Conference on Computer Vision and Pattern Recognition}}. \bibinfo{pages}{8839--8849}.
\newblock


\bibitem[Shinn et~al\mbox{.}(2023)]%
        {shinn2023reflexion}
\bibfield{author}{\bibinfo{person}{Noah Shinn}, \bibinfo{person}{Federico Cassano}, \bibinfo{person}{Ashwin Gopinath}, \bibinfo{person}{Karthik Narasimhan}, {and} \bibinfo{person}{Shunyu Yao}.} \bibinfo{year}{2023}\natexlab{}.
\newblock \showarticletitle{Reflexion: Language Agents with Verbal Reinforcement Learning}. In \bibinfo{booktitle}{\emph{NeurIPS}}.
\newblock
\urldef\tempurl%
\url{https://papers.nips.cc/paper_files/paper/2023/hash/1b44b878bb782e6954cd888628510e90-Abstract-Conference.html}
\showURL{%
\tempurl}


\bibitem[Siddiqui et~al\mbox{.}(2025)]%
        {siddiqui2025script}
\bibfield{author}{\bibinfo{person}{Momin~N Siddiqui}, \bibinfo{person}{Roy~D Pea}, {and} \bibinfo{person}{Hari Subramonyam}.} \bibinfo{year}{2025}\natexlab{}.
\newblock \showarticletitle{Script\&Shift: A layered interface paradigm for integrating content development and rhetorical strategy with llm writing assistants}. In \bibinfo{booktitle}{\emph{Proceedings of the 2025 CHI Conference on Human Factors in Computing Systems}}. \bibinfo{pages}{1--19}.
\newblock


\bibitem[Spitz(2012)]%
        {spitz2012beatles}
\bibfield{author}{\bibinfo{person}{Bob Spitz}.} \bibinfo{year}{2012}\natexlab{}.
\newblock \bibinfo{booktitle}{\emph{The Beatles: the biography}}.
\newblock \bibinfo{publisher}{Hachette UK}.
\newblock


\bibitem[Suchman(2007)]%
        {suchman2007human}
\bibfield{author}{\bibinfo{person}{Lucille~Alice Suchman}.} \bibinfo{year}{2007}\natexlab{}.
\newblock \bibinfo{booktitle}{\emph{Human-machine reconfigurations: Plans and situated actions}}.
\newblock \bibinfo{publisher}{Cambridge university press}.
\newblock


\bibitem[{Suno}(2025)]%
        {suno2024}
\bibfield{author}{\bibinfo{person}{{Suno}}.} \bibinfo{year}{2025}\natexlab{}.
\newblock \bibinfo{title}{Official website}.
\newblock \bibinfo{howpublished}{\url{https://suno.com}}.
\newblock
\newblock
\shownote{Accessed: 2025-08-01}.


\bibitem[Tanksley et~al\mbox{.}(2025)]%
        {tanksley2025ethics}
\bibfield{author}{\bibinfo{person}{Tiera Tanksley}, \bibinfo{person}{Angela~DR Smith}, \bibinfo{person}{Saloni Sharma}, {and} \bibinfo{person}{Earl~W Huff~Jr}.} \bibinfo{year}{2025}\natexlab{}.
\newblock \showarticletitle{" Ethics is not neutral": Understanding Ethical and Responsible AI Design from the Lenses of Black Youth}. In \bibinfo{booktitle}{\emph{Proceedings of the 2025 CHI Conference on Human Factors in Computing Systems}}. \bibinfo{pages}{1--20}.
\newblock


\bibitem[Vaccaro et~al\mbox{.}(2019)]%
        {vaccaro2019contestability}
\bibfield{author}{\bibinfo{person}{Kristen Vaccaro}, \bibinfo{person}{Karrie Karahalios}, \bibinfo{person}{Deirdre~K Mulligan}, \bibinfo{person}{Daniel Kluttz}, {and} \bibinfo{person}{Tad Hirsch}.} \bibinfo{year}{2019}\natexlab{}.
\newblock \showarticletitle{Contestability in algorithmic systems}. In \bibinfo{booktitle}{\emph{Companion Publication of the 2019 Conference on Computer Supported Cooperative Work and Social Computing}}. \bibinfo{pages}{523--527}.
\newblock


\bibitem[Wan et~al\mbox{.}(2024)]%
        {wan2024felt}
\bibfield{author}{\bibinfo{person}{Qian Wan}, \bibinfo{person}{Siying Hu}, \bibinfo{person}{Yu Zhang}, \bibinfo{person}{Piaohong Wang}, \bibinfo{person}{Bo Wen}, {and} \bibinfo{person}{Zhicong Lu}.} \bibinfo{year}{2024}\natexlab{}.
\newblock \showarticletitle{" It Felt Like Having a Second Mind": Investigating Human-AI Co-creativity in Prewriting with Large Language Models}.
\newblock \bibinfo{journal}{\emph{Proceedings of the ACM on human-computer interaction}} \bibinfo{volume}{8}, \bibinfo{number}{CSCW1} (\bibinfo{year}{2024}), \bibinfo{pages}{1--26}.
\newblock


\bibitem[Wang et~al\mbox{.}(2025)]%
        {Wang2025}
\bibfield{author}{\bibinfo{person}{Wen-Fan Wang}, \bibinfo{person}{Chien-Ting Lu}, \bibinfo{person}{Nil~Ponsa Campany{\`a}}, \bibinfo{person}{Bing-Yu Chen}, {and} \bibinfo{person}{Mike~Y. Chen}.} \bibinfo{year}{2025}\natexlab{}.
\newblock \showarticletitle{{AIdeation}: Designing a Human–AI Collaborative Ideation System for Concept Designers}. In \bibinfo{booktitle}{\emph{Proceedings of the 2025 CHI Conference on Human Factors in Computing Systems}} \emph{(\bibinfo{series}{CHI '25})}. \bibinfo{publisher}{ACM}.
\newblock
\href{https://doi.org/10.1145/3706598.3714148}{doi:\nolinkurl{10.1145/3706598.3714148}}


\bibitem[Wang et~al\mbox{.}(2023)]%
        {wang2023selfconsistency}
\bibfield{author}{\bibinfo{person}{Xuezhi Wang}, \bibinfo{person}{Jason Wei}, \bibinfo{person}{Dale Schuurmans}, {and} \bibinfo{person}{et al.}} \bibinfo{year}{2023}\natexlab{}.
\newblock \showarticletitle{Self-Consistency Improves Chain of Thought Reasoning in Language Models}. In \bibinfo{booktitle}{\emph{ICLR}}.
\newblock
\urldef\tempurl%
\url{https://openreview.net/forum?id=1PL1NIMMrw}
\showURL{%
\tempurl}


\bibitem[Wang et~al\mbox{.}(2024)]%
        {Wang2024}
\bibfield{author}{\bibinfo{person}{Zhijie Wang}, \bibinfo{person}{Yuheng Huang}, \bibinfo{person}{Da Song}, \bibinfo{person}{Lei Ma}, {and} \bibinfo{person}{Tianyi Zhang}.} \bibinfo{year}{2024}\natexlab{}.
\newblock \showarticletitle{PromptCharm: Text-to-Image Generation through Multi-modal Prompting and Refinement}. In \bibinfo{booktitle}{\emph{Proceedings of the 2024 CHI Conference on Human Factors in Computing Systems}} \emph{(\bibinfo{series}{CHI '24})}. \bibinfo{publisher}{ACM}, \bibinfo{pages}{185:1--185:21}.
\newblock
\href{https://doi.org/10.1145/3613904.3642803}{doi:\nolinkurl{10.1145/3613904.3642803}}


\bibitem[Wei et~al\mbox{.}(2022)]%
        {wei2022chain}
\bibfield{author}{\bibinfo{person}{Jason Wei}, \bibinfo{person}{Xuezhi Wang}, \bibinfo{person}{Dale Schuurmans}, {and} \bibinfo{person}{et al.}} \bibinfo{year}{2022}\natexlab{}.
\newblock \showarticletitle{Chain-of-Thought Prompting Elicits Reasoning in Large Language Models}. In \bibinfo{booktitle}{\emph{NeurIPS}}.
\newblock
\urldef\tempurl%
\url{https://proceedings.neurips.cc/paper_files/paper/2022/hash/9d560961f0aa0a8a1e3953eacb5f63b0-Abstract-Conference.html}
\showURL{%
\tempurl}


\bibitem[Wittgenstein(1966)]%
        {wittgenstein1966lectures}
\bibfield{author}{\bibinfo{person}{Ludwig Wittgenstein}.} \bibinfo{year}{1966}\natexlab{}.
\newblock \bibinfo{booktitle}{\emph{Lectures \& conversations on aesthetics, psychology, and religious belief}}.
\newblock \bibinfo{publisher}{Univ of California Press}.
\newblock


\bibitem[Xu et~al\mbox{.}(2024)]%
        {xu2024symbcot}
\bibfield{author}{\bibinfo{person}{Jundong Xu}, \bibinfo{person}{Hao Fei}, \bibinfo{person}{Liangming Pan}, \bibinfo{person}{Qian Liu}, \bibinfo{person}{Mong-Li Lee}, {and} \bibinfo{person}{Wynne Hsu}.} \bibinfo{year}{2024}\natexlab{}.
\newblock \showarticletitle{Faithful Logical Reasoning via Symbolic Chain-of-Thought}. In \bibinfo{booktitle}{\emph{ACL}}.
\newblock
\urldef\tempurl%
\url{https://aclanthology.org/2024.acl-long.720/}
\showURL{%
\tempurl}


\bibitem[Yao et~al\mbox{.}(2023a)]%
        {yao2023react}
\bibfield{author}{\bibinfo{person}{Shunyu Yao}, \bibinfo{person}{Jeffrey Zhao}, \bibinfo{person}{Dian Yu}, {and} \bibinfo{person}{et al.}} \bibinfo{year}{2023}\natexlab{a}.
\newblock \showarticletitle{{ReAct}: Synergizing Reasoning and Acting in Language Models}. In \bibinfo{booktitle}{\emph{ICLR}}.
\newblock
\urldef\tempurl%
\url{https://arxiv.org/abs/2210.03629}
\showURL{%
\tempurl}


\bibitem[Yao et~al\mbox{.}(2023b)]%
        {yao2023treeofthoughts}
\bibfield{author}{\bibinfo{person}{Shunyu Yao}, \bibinfo{person}{Jeffrey Zhao}, \bibinfo{person}{Dian Yu}, {and} \bibinfo{person}{et al.}} \bibinfo{year}{2023}\natexlab{b}.
\newblock \showarticletitle{Tree of Thoughts: Deliberate Problem Solving with Large Language Models}. In \bibinfo{booktitle}{\emph{NeurIPS}}.
\newblock
\urldef\tempurl%
\url{https://arxiv.org/abs/2305.10601}
\showURL{%
\tempurl}


\bibitem[Yurrita et~al\mbox{.}(2023)]%
        {yurrita2023disentangling}
\bibfield{author}{\bibinfo{person}{Mireia Yurrita}, \bibinfo{person}{Tim Draws}, \bibinfo{person}{Agathe Balayn}, \bibinfo{person}{Dave Murray-Rust}, \bibinfo{person}{Nava Tintarev}, {and} \bibinfo{person}{Alessandro Bozzon}.} \bibinfo{year}{2023}\natexlab{}.
\newblock \showarticletitle{Disentangling fairness perceptions in algorithmic decision-making: the effects of explanations, human oversight, and contestability}. In \bibinfo{booktitle}{\emph{Proceedings of the 2023 CHI Conference on Human Factors in Computing Systems}}. \bibinfo{pages}{1--21}.
\newblock


\bibitem[Zeng et~al\mbox{.}(2024a)]%
        {Zeng2024}
\bibfield{author}{\bibinfo{person}{Xingchen Zeng}, \bibinfo{person}{Ziyao Gao}, \bibinfo{person}{Yilin Ye}, {and} \bibinfo{person}{Wei Zeng}.} \bibinfo{year}{2024}\natexlab{a}.
\newblock \showarticletitle{IntentTuner: An Interactive Framework for Integrating Human Intentions in Fine-Tuning Text-to-Image Generative Models}. In \bibinfo{booktitle}{\emph{Proceedings of the 2024 {CHI} Conference on Human Factors in Computing Systems}} \emph{(\bibinfo{series}{CHI '24})}. \bibinfo{publisher}{ACM}.
\newblock
\href{https://doi.org/10.1145/3613904.3642165}{doi:\nolinkurl{10.1145/3613904.3642165}}


\bibitem[Zeng et~al\mbox{.}(2024b)]%
        {IntentTuner}
\bibfield{author}{\bibinfo{person}{Xingchen Zeng}, \bibinfo{person}{Ziyao Gao}, \bibinfo{person}{Yilin Ye}, {and} \bibinfo{person}{Wei Zeng}.} \bibinfo{year}{2024}\natexlab{b}.
\newblock \showarticletitle{IntentTuner: An Interactive Framework for Integrating Human Intentions in Fine-tuning Text-to-Image Generative Models}. In \bibinfo{booktitle}{\emph{Proceedings of the 2024 CHI Conference on Human Factors in Computing Systems (CHI ’24)}}. \bibinfo{publisher}{ACM}.
\newblock
\href{https://doi.org/10.1145/3613904.3642165}{doi:\nolinkurl{10.1145/3613904.3642165}}


\bibitem[Zhang et~al\mbox{.}(2022)]%
        {zhang2022storydrawer}
\bibfield{author}{\bibinfo{person}{Chao Zhang}, \bibinfo{person}{Cheng Yao}, \bibinfo{person}{Jiayi Wu}, \bibinfo{person}{Weijia Lin}, \bibinfo{person}{Lijuan Liu}, \bibinfo{person}{Ge Yan}, {and} \bibinfo{person}{Fangtian Ying}.} \bibinfo{year}{2022}\natexlab{}.
\newblock \showarticletitle{StoryDrawer: a child--AI collaborative drawing system to support children's creative visual storytelling}. In \bibinfo{booktitle}{\emph{Proceedings of the 2022 CHI conference on human factors in computing systems}}. \bibinfo{pages}{1--15}.
\newblock


\bibitem[Zhang et~al\mbox{.}(2024)]%
        {zhang2024improvingcot}
\bibfield{author}{\bibinfo{person}{Xuan Zhang}, \bibinfo{person}{Chao Du}, \bibinfo{person}{Tianyu Pang}, {and} \bibinfo{person}{et al.}} \bibinfo{year}{2024}\natexlab{}.
\newblock \showarticletitle{Improving Chain-of-Thought Reasoning in LLMs via Chain-of-Preference Optimization}. In \bibinfo{booktitle}{\emph{NeurIPS}}.
\newblock
\urldef\tempurl%
\url{https://proceedings.neurips.cc/paper_files/paper/2024/file/00d80722b756de0166523a87805dd00f-Paper-Conference.pdf}
\showURL{%
\tempurl}


\bibitem[Zhao et~al\mbox{.}(2023)]%
        {zhao2023verifyandedit}
\bibfield{author}{\bibinfo{person}{Ruochen Zhao}, \bibinfo{person}{Xingxuan Li}, \bibinfo{person}{Shafiq Joty}, \bibinfo{person}{Chengwei Qin}, {and} \bibinfo{person}{Lidong Bing}.} \bibinfo{year}{2023}\natexlab{}.
\newblock \showarticletitle{Verify-and-Edit: A Knowledge-Enhanced Chain-of-Thought Framework}. In \bibinfo{booktitle}{\emph{ACL}}.
\newblock
\urldef\tempurl%
\url{https://aclanthology.org/2023.acl-long.320/}
\showURL{%
\tempurl}


\bibitem[Zheng et~al\mbox{.}(2025)]%
        {zheng2025customizing}
\bibfield{author}{\bibinfo{person}{Xi Zheng}, \bibinfo{person}{Zhuoyang Li}, \bibinfo{person}{Xinning Gui}, {and} \bibinfo{person}{Yuhan Luo}.} \bibinfo{year}{2025}\natexlab{}.
\newblock \showarticletitle{Customizing emotional support: How do individuals construct and interact with LLM-powered chatbots}. In \bibinfo{booktitle}{\emph{Proceedings of the 2025 CHI Conference on Human Factors in Computing Systems}}. \bibinfo{pages}{1--20}.
\newblock


\bibitem[Zhong et~al\mbox{.}(2024a)]%
        {zhong2024ai}
\bibfield{author}{\bibinfo{person}{Ruican Zhong}, \bibinfo{person}{Donghoon Shin}, \bibinfo{person}{Rosemary Meza}, \bibinfo{person}{Predrag Klasnja}, \bibinfo{person}{Lucas Colusso}, {and} \bibinfo{person}{Gary Hsieh}.} \bibinfo{year}{2024}\natexlab{a}.
\newblock \showarticletitle{AI-assisted causal pathway diagram for human-centered design}. In \bibinfo{booktitle}{\emph{Proceedings of the 2024 CHI Conference on Human Factors in Computing Systems}}. \bibinfo{pages}{1--19}.
\newblock


\bibitem[Zhong et~al\mbox{.}(2024b)]%
        {Zhong2024}
\bibfield{author}{\bibinfo{person}{Ruican Zhong}, \bibinfo{person}{Donghoon Shin}, \bibinfo{person}{Rosemary Meza}, \bibinfo{person}{Predrag Klasnja}, \bibinfo{person}{Lucas Colusso}, {and} \bibinfo{person}{Gary Hsieh}.} \bibinfo{year}{2024}\natexlab{b}.
\newblock \showarticletitle{AI-Assisted Causal Pathway Diagram for Human-Centered Design}. In \bibinfo{booktitle}{\emph{Proceedings of the 2024 {CHI} Conference on Human Factors in Computing Systems}} \emph{(\bibinfo{series}{CHI '24})}. \bibinfo{publisher}{ACM}.
\newblock
\href{https://doi.org/10.1145/3613904.3642179}{doi:\nolinkurl{10.1145/3613904.3642179}}


\bibitem[Zhou et~al\mbox{.}(2023)]%
        {zhou2023leasttomost}
\bibfield{author}{\bibinfo{person}{Yao Zhou}, \bibinfo{person}{Nathanael Sch{\"a}rli}, \bibinfo{person}{Le Hou}, {and} \bibinfo{person}{et al.}} \bibinfo{year}{2023}\natexlab{}.
\newblock \showarticletitle{Least-to-Most Prompting Enables Complex Reasoning in Large Language Models}. In \bibinfo{booktitle}{\emph{ICLR}}.
\newblock
\urldef\tempurl%
\url{https://openreview.net/forum?id=WZH7099tgf}
\showURL{%
\tempurl}


\end{thebibliography}

\appendix

\end{document}